\documentclass[12pt]{article}
\usepackage[utf8]{inputenc}
\usepackage[round]{natbib}
\usepackage{tikz}
\usepackage[colorlinks=true,urlcolor=blue,citecolor=magenta]{hyperref}
\usepackage{amssymb,amsmath,amsfonts}
\usepackage{epsfig}
\usepackage{epsfig}
\usepackage{graphicx}
\usepackage{epstopdf}
\usepackage{caption}
\usepackage{subcaption}
\usepackage{amscd}
\usepackage{amsthm}
\usepackage{latexsym}
\usepackage{float}
\usepackage{mathtools}
\usepackage{amsbsy}
\usepackage{bbm}
\usepackage[english]{babel}
\usepackage{psfrag}
\usepackage{tabularx}
\usepackage{colortbl}
\usepackage{xcolor}
\usepackage{braket}
\usepackage{csquotes}
\usepackage[stable]{footmisc}
\usepackage{setspace}
\newcommand{\bea}{\begin{eqnarray}}
\newcommand{\eea}{\end{eqnarray}}
\newcommand{\eq}[1]{(\ref{#1})}
\def\nn{\nonumber\\}
\def\na{\nabla}
\def\half{\frac{1}{2}\,}
\def\Tr{{\rm Tr}}
\def\dd{\mbox{d}}

\def\bra{\langle}
\def\ket{\rangle}

\def\b{\beta}

\def\e{\epsilon}

\def\f{\phi}

\def\l{\lambda}

\def\m{\mu}
\def\n{\nu}
\def\s{\sigma}

\def\pa{\partial}
\newcommand{\ti}[1]{\tilde{#1}}

\newcommand{\sm}[1]{\mbox{\scriptsize #1}}
\newcommand{\tn}[1]{\mbox{\tiny #1}}

\makeatletter
\def\blfootnote{\xdef\@thefnmark{*}\@footnotetext}
\makeatother

\usepackage{authblk}
\usepackage{lmodern}

\title{Dualities in Physics\blfootnote{This is the pre-submission version of a Cambridge Element in the Foundations of Contemporary Physics. The book is published by Cambridge University Press and is available here:
https://www.cambridge.org/core/elements/dualities-in-physics/45667B3C71A2C7A5B347748239243EC7. When citing this work, please refer to the published version.}}

\author[1]{Sebastian De Haro}
\author[2]{Enrico Cinti}
\affil[1]{Institute for Logic, Language and Computation and Institute of Physics, University of Amsterdam}
\affil[2]{Department of Philosophy, University of Geneva} %, 5, rue de Candolle, CH-1211 Gen\`eve 4}
\date{7 May 2026}

\begin{document}

\pagenumbering{gobble}

\maketitle

\newpage

\begin{abstract}

For more than half a century, dualities have been at the heart of modern physics. From quantum mechanics to statistical mechanics, condensed matter physics, quantum field theory and quantum gravity, dualities have proven useful in solving problems that are otherwise quite intractable. Being surprising and unexpected, dualities have been taken to raise philosophical questions about the nature and formulation of scientific theories, scientific realism, emergence, symmetries, explanation, understanding, and theory construction. This book discusses what dualities are, gives a selection of examples, explores the themes and roles that make dualities interesting, and highlights their most salient types. It aims to be an entry point into discussions of dualities in both physics and philosophy. The philosophical discussion emphasises three main topics: whether duals are theoretically equivalent, the view of scientific theories that is suggested by dualities (namely, a geometric view of theories), and the compatibility between duality and emergence.\\\\%For readers who are interested in quick answers to philosophical questions, the book includes a number of proposed FAQs.
    
\end{abstract}

%\centering\copyright~ Sebastian De Haro, Enrico Cinti, 2025

\newpage

\tableofcontents

\newpage

\pagenumbering{arabic}

\section{Introduction}

For more than half a century, dualities have been at the heart of modern physics. From statistical mechanics to quantum gravity, dualities are frequently used to solve problems that would otherwise be quite intractable. Indeed, dualities are valuable tools for understanding physical mechanisms, constructing new theories, and developing novel interpretations. Thus an understanding of dualities is indispensable for engaging with key aspects of both past and current developments in physics.

Dualities are not a special topic in high-energy physics, nor are they purely formal properties of speculative quantum gravity theories. For we encounter dualities in a wide variety of physical theories, from quantum mechanics to statistical mechanics, from electrodynamics to condensed matter physics, and indeed in quantum field theory and quantum gravity.

But what is a duality? A duality is an equivalence between theories: as we will discuss, it is an appropriate {\it isomorphism}. Being equivalences between scientific theories, it is natural for dualities to be of interest to both physics and philosophy. For both fields are, each for its own reasons and from its perspective, concerned with the formulation, interpretation and equivalence of physical theories.

We will argue that dualities are a natural setting for discussions of some of philosophy's traditional questions: about the nature and formulation of scientific theories, the nature of explanation and understanding, scientific realism, emergence (especially the emergence of spacetime), theoretical equivalence, symmetries, and the heuristics of theory construction. 

Furthermore, since the debates about these topics are ongoing, the philosophical discussion of dualities is not a matter of taking agreed notions of, say, explanation or emergence, and illustrating them for dual theories: rather, it is a matter of letting dualities interact with these debates in a fruitful way. We here briefly illustrate what we mean by `fruitful interaction' for two specific debates: more details follow in later Chapters.

First, philosophers of dualities have contributed to the debate about theoretical equivalence, i.e.~what it means for two theories to `say the same thing, in different words', which has been a major topic of discussion in the philosophy of dualities in recent years. Most philosophers have advocated a mixed view of theoretical equivalence, i.e.~one that requires both formal and (substantive) interpretative criteria. And, although the details differ, all hands agree that the question {\it Are duals theoretically equivalent?} has no automatic yes or no answer. The duality-based analyses of theoretical equivalence that have emerged from these discussions bear on the question of theoretical equivalence more generally.\footnote{See the references in Chapter \ref{philiss} (Section \ref{TE}), where we will also advocate a mixed criterion of theoretical equivalence, with duality being the formal part of the criterion.}

Second, dualities bear on the debate over the best formulation of scientific theories, i.e.~on the question {\it What is a scientific theory?} Perhaps surprisingly, dualities (and their generalisations, quasi-dualities) suggest that the standard semantic conception, according to which a scientific theory is a collection of models, is insufficient to describe theories in physics, because (i) it does not take into account the fact that the set of models is often equipped with {\it geometric structure},\footnote{Note that our use of 
`geometric structure' includes algebraic-geometric structures, i.e.~we adopt a broad and comprehensive view of what counts as geometric structure. The relevant contrast is with set-theoretic structure, rather than with non-geometric mathematical structures.} and (ii) it is silent about the physical role and interpretation of such structure. 

Indeed, dualities suggest exactly this possibility: so that, in some key cases, a theory is best formulated as a differentiable manifold, rather than as a set.\footnote{For a discussion of this second question, see Section \ref{QD}.} 
Following \cite{deharobutterfieldOUP}, we will dub this the {\it geometric view of theories}. An important feature of this discussion is that it illustrates the advantage of looking at examples of dual theories from a more systematic perspective, thereby making fruitful connections with parallel discussions in e.g.~logic-oriented philosophy of science.

This book offers a discussion of what dualities are, a selection of examples and a discussion of selected philosophical questions that they raise. The book is aimed at readers with an interest in examples of dualities, their relation to current issues in physics, and the philosophy of dualities. Thus we expect both readers who are physicists, and readers who are philosophers, to benefit from reading this book. (For more details and in-depth discussion we will, in various places, refer to the literature.)

In the rest of this Introduction, Section \ref{Schemaex} first briefly motivates our specific treatment of dualities, which we call a Schema, and the significance of the examples that we have chosen. Section \ref{features} then discusses some of the main features of dualities that make them both physically and philosophically interesting.

\subsection{A Schema for dualities, and examples}\label{Schemaex}

In this book, our usage of the word `duality' will be the standard one in physics: a duality is an {\it isomorphism between physical theories}.\footnote{There is a vast literature on dualities, and so we will here only cite a sample of papers in physics that discuss a variety of examples of dualities as being isomorphisms: \citeauthor{aharony2000large} (2000:~pp.~190, 254), \citeauthor{polchinski2} (1998:~pp.~402, 404, 425), \citeauthor{becker2007string} (2007:~p.~412, 675) and \citeauthor{dijkgraaf1997houches} (1997:~p.~143).} 
This conception of a duality, applied to the several specific forms that a physical theory normally takes (for example, a theory viewed in terms of states, quantities and dynamics), is what \citet{de2018schema} call a Schema for dualities.  We will discuss the Schema in more detail in Section \ref{dualint}. For now, however, let us mention that we will adopt this usage of the Schema in the same undogmatic spirit with which De Haro and Butterfield have proposed it: namely, they accept that there may be ``rough edges'' in matching it to physicists' examples of dualities.\footnote{In what follows, we will in general distinguish dualities and {\it quasi-dualities}, which include, as a particularly important subcase, effective dualities (and this phrase is commonly used in physics). While some physicists do not make this distinction, it is important, in order to avoid confusion, to disambiguate the use of the word `duality' in some such way.} 
Furthermore, like De Haro and Butterfield, we accept that there may be more mathematically precise conceptions of duality than one that uses the notion of isomorphism.\footnote{Thus we very much welcome reformulations of dualities that use category theory or other mathematical tools, provided that they can cover the full range of detailed examples in physics that the Schema covers. However, we think that some of the recent criticisms of the application of isomorphism criteria in physics are unsuccessful: see the discussion in \citeauthor{deharobutterfieldOUP} (2025:~Chapter 11).} 

With that said, we agree that the Schema fares very well, since it describes both classic (and elementary), as well as technically more demanding, examples of dualities in physics. Furthermore, as we will demonstrate in Chapter \ref{philiss}, it enables a precise discussion of the relevant philosophical questions, and it casts light on various issues.\\
\\
{\bf Choice of examples:} our choice has been guided by our aim of giving an even treatment along the following three axes: 

\begin{itemize}
    \item[(i)] familiar and less familiar examples,
    \item[(ii)] elementary and more technically demanding examples, and  
    \item[(iii)] examples that are representative of a variety of branches of physics: quantum mechanics, statistical mechanics, electrodynamics, condensed matter physics, quantum field theory and quantum gravity.  
\end{itemize} 
Thus, for example, we include a discussion of Kramers-Wannier duality (Section \ref{KWd}) for a two-dimensional Ising lattice. Although this duality is elementary, it has received very little attention in the philosophical literature on dualities, and so it will be less familiar to philosophers.\footnote{The only discussion of Kramers-Wannier duality in the philosophical literature that we are aware of is the brief discussion by Polchinski: see \citeauthor{polchinski2017dualities} (2017:~p.~8). See also \citeauthor{deharobutterfieldOUP} (2025:~Section 4.4).} However, Kramers-Wannier duality is in some sense a paradigmatic example of the kind of physical phenomena that are associated with dualities, for it allows the precise study of the phase transition between a ferromagnetic and a paramagnetic phase, symmetry breaking, and the emergence of an ordered state of matter.

Other dualities that build on this elementary example, and that show similar behaviour, are particle-soliton dualities in three dimensions (two dimensions of space and one of time: see Section \ref{Psd3D}), which are associated with the Berezinskii-Kosterlitz-Thouless phase transition between a high-temperature phase of dissociated vortices and a low-temperature phase of condensated pairs of vortices and anti-vortices.

Point (iii) leads to some unexpected connections. For example, there are (under the umbrella of AdS-CFT) surprising dualities that map solid state systems to (quantum) gravity systems (\citeauthor{mauri2024gauge}, 2024). This is surprising, because, apart from the fact that the solid state systems in question are actually realized in the lab (\citeauthor{smit2021momentum}, 2021), the duality relation goes well beyond the analogies that have been discussed in the philosophical literature:\footnote{See \citeauthor{fraser2016higgs} 2016; \citeauthor{dardashti2017confirmation} 2017; \citeauthor{dardashti2019hawking} 2019; \citeauthor{fraser2020development} 2020; \citeauthor{crowther2021we} 2021.} in particular, it provides a much closer connection between condensed matter physics and high-energy physics than previously considered examples.\footnote{While we will not undertake a detailed philosophical exploration of this duality here, it is clearly a fruitful avenue for research. See \citet{cinti2025holographic}
}

In the philosophical literature, inter-theoretic relations such as theoretical equivalence, reduction and emergence have traditionally been discussed between similar, or at least closely related, branches of physics. Standard examples of theoretical equivalence are the relations between different versions of the Maxwell theory, between different versions of Newtonian mechanics, and between different versions of general relativity; and standard examples of reduction and emergence are the relations between statistical mechanics and thermodynamics and between high-energy and low-energy quantum field theories (arguably, the emergence of classical mechanics from quantum mechanics is one example where the distances are greater). Indeed, the discussion of inter-theoretic relations between more distant branches of physics, where the interpretations of the two theories differ more, has often been mostly in discussions of formal analogies, e.g.~between condensed matter physics and quantum field theory.\footnote{See, for example, \citeauthor{fraser2016higgs} (2016:~p.~72).} What strikes us as novel about the examples of dualities, such as the example of condensed matter and AdS-CFT above, is that these relations are not formal analogies but isomorphisms (or close to isomorphisms): and furthermore that, even if the physics on the two sides of a duality sometimes looks very different, there is nevertheless a deep match of the physics across the duality, i.e.~not only of formulas and formalisms, but also of detailed physical mechanisms, which can be used to answer questions about, and even to (re)construct, the mechanisms in the dual model. This ``matching of physical mechanisms'', and the related ideas of modelling, simulation and use of analogies across dualities, surely deserve further philosophical analysis.

\subsection{Features of Dualities}\label{features}

There are three main features of dualities that make them particularly salient in both physics and philosophy: dualities are often surprising, they involve rich physics, and they are useful in theory construction (i.e.~they have heuristic power). In this Section, we will discuss how dualities are surprising and useful: it will be the task of the following Chapters to illustrate the physical richness of the examples.

As we mentioned when we discussed our choice of examples (see point (iii) above), dualities are often {\bf surprising}, because they relate very different-looking theories, often across different areas of physics. For example, particle-soliton dualities usually map different phases of matter onto each other: a phase with topological order is mapped onto a disordered phase. Indeed, these types of dualities normally exchange: (a) a conserved {\it topological} current, which is not derived from a symmetry, and whose conservation follows from the topology of the system; with (b) a Noether current that has an associated symmetry. 

A particularly salient type of duality is {\it quantum duality}, where two dual formulations of a single quantum theory are obtained from the quantization of two classical theories that are inequivalent to each other. This is the case in AdS-CFT, where a quantum gravity theory in AdS is believed to be dual to a conformal quantum field theory. At the quantum level, the duality often amounts to a unitary transformation, but the classical limits are strikingly different, which means that a single quantum theory has two very different limits where it becomes classical. 

To illustrate the {\bf usefulness}, in the sense of the {\bf heuristic power}, of dualities, note that dualities that were originally discovered for condensed matter systems like superconductors, were later used in an influential proposal of 't Hooft and Mandelstam for a mechanism of colour charge confinement, which is still a major open problem in the theory of quantum chromodynamics. Also, recent developments in quantum field theory and string theory cannot be understood without the examples of dualities in statistical mechanics and condensed matter theory (especially the particle-soliton dualities discussed above), because many of the modern ideas about key behaviours in quantum field theory 
(symmetry breaking, phase transitions, colour confinement, etc.) grew out of the use of dualities and analogies with condensed matter systems. Likewise, the progress in understanding the non-perturbative properties of string theory (especially black hole entropy counting, the behaviour of D-branes, and M-theory) is best understood in connection with the key behaviours in quantum field theory that we just mentioned. Thus, although the philosophical literature has often focussed on dualities in string theory, dualities in quantum field theory, condensed matter physics and statistical mechanics deserve further study. \\
\\
{\bf Preliminary note and prospectus.} This book draws from \citeauthor{deharobutterfieldOUP} (2025) in: (a) its general approach to dualities (namely, the introduction of a Schema for dualities), (b) its choice of examples, and (c) its philosophical discussions. For this reason, we will not always refer back to \citeauthor{deharobutterfieldOUP} (2025), except when we use specific results. But the reader can rest assured that, on all the topics (except for the role of emergence in bulk reconstruction, see Sections \ref{sec: bulkads} and \ref{sec: emerg}), more detailed results and explanations can be found there. 

Having said that, the present book also contains novel aspects, and aspects that are presented slightly differently from how they are discussed in \citeauthor{deharobutterfieldOUP} (2025). For example, the discussion, at the end of Section \ref{EMMa}, of the common core theory for the simple electric-magnetic duality of the Maxwell theory in vacuum, differs from the one in \citeauthor{deharobutterfieldOUP} (2025) in that it is formulated for the Maxwell theory written in components, rather than for the Lorentz-invariant theory. Also, the discussion, in Section \ref{sec: bulkads}, of bulk reconstruction in AdS-CFT, and its relation to emergence discussed in Section \ref{sec: emerg}, is wholly new. Finally, Section \ref{sec: faq} contains a number of FAQs about dualities that may be convenient for readers looking for quick answers to specific conceptual questions about dualities.

The plan of the rest of the book is as follows. Chapter \ref{dualint} introduces the notion of duality and its main features. Chapters \ref{3} and \ref{DQFTG} are a brief introduction to a selection of examples of dualities: Chapter \ref{3} introduces classic, more elementary, examples, and Chapter \ref{DQFTG} gives physically more advanced examples. Then Chapter \ref{philiss} discusses a selection of philosophical questions associated with dualities, and also includes answers to a number of FAQs. Chapter \ref{conclusion} concludes.

\section{Dualities and their Roles}\label{dualint}

As the Introduction emphasised, dualities are, in both philosophy and physics, a rich and timely topic: indeed ideal for the kind of interdisciplinary project that we envision philosophy of physics to be. Thus the hallmark of such a project, which we hope to illustrate in the rest of this book, is the interaction between the technical issues in the physics of dualities and their philosophical interpretation, and how this interaction can fruitfully enrich both fields, which in turn enhances our understanding of dualities.

Also, it pays off, before delving into the technical details of various examples, to discuss dualities from a more general perspective, so as to ground our overall analysis in the later Chapters. This task is the goal of this Chapter.

Section \ref{2.1} will give a general characterization of the structure of duality relations, and of the kinds of interpretative options that are available when facing two dual theories. In particular, we will focus on the conception of duality as an \textit{isomorphism}, especially as in the Schema proposed in \citeauthor{de2018schema} (2018) and \citeauthor{deharobutterfieldOUP} (2025). We will also discuss the contrast between internal and external interpretations, i.e.~between an interpretation where we only commit to duality-invariant facts, and an interpretation where we also commit to facts that are specific to one of the duals.

In Section \ref{2.2}, we will discuss the themes, roles and types of dualities, that we will encounter throughout this book. Notable examples of {\bf themes} are the \textit{hard-easy} character of dualities, whereby calculations that are hard in one dual theory are easy in the other, and vice versa: or \textit{elementary-composite}, whereby entities that cannot be further decomposed in one theory are mapped to composites of simple entities in the dual theory, and vice versa. 

About the {\bf roles} of dualities, an important distinction will be between the \textit{theoretical} and \textit{heuristic} roles of dualities. 
In their {\it theoretical role}, dualities are fixed formal properties of particular physical theories: in their {\it heuristic role}, dualities are guides towards the development of more fundamental theories underlying the duality.

Finally, we will look at the different {\bf types} of dualities: here, an important type will be \textit{quantum dualities}, i.e.~dualities between theories that are equivalent at the quantum, but not at the classical, level, where the two dual theories appear as different classical limits of a single quantum theory. We will also discuss \textit{quasi-dualities}, which are similar to, but fall short of being, dualities. As we will see, (quasi-)dualities are closely tied to the structure of scientific theories, and in particular they lead to a different conception of scientific theories: namely, the \textit{geometric view} of theories.

\subsection{Dualities: What They Are and How to Interpret Them}\label{2.1}

In this first Section, our goal is to introduce terminology and notation to help us understand the fundamental structure of dualities. The point of this jargon is both to understand how dualities can be defined as formal relations between theories, and how these formal relations bear on how we interpret theories. The approach that we will follow in formalizing the structure of dualities is known as the \textit{Schema} for dualities.\footnote{\citeauthor{deharobutterfieldOUP} (2025) is an exposition of this approach, which is illustrated in several dozen examples of dualities.}

We first briefly discuss a broad and intuitive characterization of dual theories, which will give us a basis to develop the Schema. By `duals', we usually mean theories that are apparently (very) different, but nonetheless isomorphic. 
%formally equivalent, where by `formal equivalence' we mean, roughly, that there is a mapping, typically an isomorphism, between the formalisms of the two theories, 
We typically understand this as an isomorphism of spaces of solutions. Whether or not this formal equivalence, i.e.~this isomorphism, translates into a claim of theoretical equivalence is a matter of dispute that need not concern us at this formal stage.\footnote{See in particular \citeauthor{butterfield2021dualities} (2021); \citeauthor{de2021theoretical} (2021) for extensive discussion of the relation between dualities and theoretical equivalence.} Note that here, by theoretical equivalence, we mean that the two theories \textit{say the same thing}, i.e.~they express the same physical content and the same physically meaningful propositions (more details in Section \ref{TE}). 

It seems natural to say that, under some interpretations, duals are theoretically equivalent, while under other interpretations they cannot be. For example, theoretical equivalence is a reasonable verdict to make when two duals are taken to describe the same physical system. But it seems a non-starter when the duals are taken to describe two distinct systems, or a single system but from two different perspectives (as in the descriptions of water given by hydrodynamics and by molecular dynamics). Nonetheless, it seems reasonable to claim that two dual theories, independent of whether or not they are theoretically equivalent, should count as empirically equivalent if they describe the same physical system with comparable predictive accuracy and success. By `empirical equivalence', we here mean that two theories agree in their observable content, i.e.~they make the same predictions for all possible observations and experiments that might be carried out on the physical systems described by them.\footnote{Note that our use of `observable' here is weaker than the standard usage in the philosophical literature, where `observable' usually means \textit{observable by the unaided senses}. However, it is important to note that on anyone's conception, empirical equivalence is an interpretative notion and not a formal i.e.~mathematical relation. Thus whether duals are empirically equivalent always depends on how they are interpreted. Empirical equivalence will not be our focus in this book. For a discussion of the relation between empirical equivalence and duality, see \citeauthor{de2020empirical} (2020, 2023) and \citet{weatherall2020equivalence}.} 

So dual theories are formally equivalent theories, and under certain conditions they can also be theoretically equivalent. But dualities are expected to obtain between theories where these sorts of equivalences would have been {\it unexpected}. \\
\\
{\bf Duality and symmetry.} It is natural to contrast duality and symmetry, understood as a map between solutions (more generally, an automorphism of state-spaces and quantities) that, at least in certain cases, is expected to leave the physical content of the theory invariant. This feature of symmetries does indeed make them apparently very similar to dualities, because both are formal equivalence maps that may allow a verdict of physical equivalence, depending on the specific physical system or theory under consideration, and on our interpretative choices regarding these equivalences. Indeed, we can think of dualities and symmetries as being related, and of a duality as a \textbf{giant symmetry} (\citeauthor{deharobutterfieldOUP} 2025). The main formal difference between the two is that dualities map whole theories, while symmetries are automorphisms of state-spaces and-or of sets of quantities (\citeauthor{caulton2015role} 2015).\\
\\
{\bf Theories and models.} The previous observation motivates the first important point to make in introducing the Schema: namely, in the presence of dualities, it is sensible to take the standard philosophical terminology of \textit{theory} and \textit{model} (which in philosophy of physics usually takes the place of the above-mentioned `solutions') and bring it \textbf{one level up}. By `one level up', we mean that \textit{model} will mean one of the ``theories'' (in the old sense) that are related by the duality map, while \textit{theory} will mean a hypothetical theory that is behind the two duals. This usage is one level up because `model' usually means solutions of theories, while `theory' means the theories themselves; on the other hand, in the case of dualities, we call `models' the theories themselves, and we call `theory' a further theory, where the duality is a manifest equivalence and which stands to the duals in a relation analogous to that between models and theories in the standard picture. More generally, and regardless of dualities, models are representations (in the mathematical not philosophical sense) or instantiations of theories.

An example will be useful to get acquainted with this jargon. During the development of quantum mechanics, two rival theories became popular to account for quantum phenomena: Heisenberg’s matrix mechanics, which described quantum phenomena using matrices and linear algebra, and Schrödinger’s wave mechanics, which described quantum phenomena in terms of waves. These two theories were considered rival attempts at formulating a theory of quantum mechanics, with various arguments given in favour of one or the other,\footnote{See \citeauthor{de2017understanding} (2017:~pp.~226-251) for a discussion of this fascinating debate. For some subtleties regarding the equivalence between matrix and wave mechanics, see \cite{muller1997equivalence}.}
until 1932, when von Neumann, in his seminal work on the mathematical foundations of quantum mechanics (\citeauthor{von1955mathematical}, 1932), showed that matrix and wave mechanics are two equivalent ways of presenting the same physical content, now expressed in terms of Hilbert spaces. In this example, the models are the two ``theories'' (old sense!) of wave and matrix mechanics, which are models of the underlying theory of quantum mechanics expressed in terms of operators on Hilbert spaces of states, and where the duality between wave and matrix mechanics is the isomorphism between the corresponding Hilbert spaces, with their respective sets of operators.\\
\\
{\bf Theories as triples of states, quantities and dynamics.} With our usage of `theory' and `model' clarified, we now introduce a more formal, but otherwise standard, definition of what a scientific theory or model minimally amounts to, from a mathematical point of view (and for the moment, regardless of physical interpretation):\footnote{Nevertheless, the perspective that we adopt in thus presenting a theory as a triple is not ``purely formal'', because the bare theory both constrains the theory's descriptive capacities, and is constrained by the kind of interpretation that one envisages for the formal triple. Thus, for example, the choice of variables that one uses to describe the state-space is often motivated by one's intended interpretation: indeed, our use of terms like `state' and `dynamics' (and other standard phrases in mathematical physics like `field', `symmetry', etc.) already indicates that a triple has a minimal interpretation, sometimes also called a `proto-interpretation'. Thus a triple and its interpretation are two sides of the same coin (namely, the physical theory), rather than two separate, pre-existing, ingredients. Of course, this does not prevent us from conceptually distinguishing them. Namely, a triple is an entity in mathematical physics, rather than in pure mathematics.} 
a theory or model is a triple $\bra\mathcal{S},\mathcal{Q},\mathcal{D}\ket$ of set of states $\mathcal{S}$ or {\it state-space}, set of quantities $\mathcal{Q}$, and dynamics $\mathcal{D}$. `Set of states' here means, broadly, the space of variables that, once interpreted, will describe the possible configurations of a physical system or sets of systems that we are interested in, formally represented in the triple by things like, e.g.~points in phase space or rays in Hilbert space. `Quantities' are the variables that, once interpreted, will describe the observable physical properties of the system: in the triple, these properties are formally represented by, for example, linear operators (in a quantum theory) or, in classical mechanics, by functions on phase space. Finally, `dynamics' is some set of equations that, once interpreted, describe the evolution of the system we are studying, such as the Schrödinger equation in quantum mechanics or the Einstein field equations in General Relativity.\footnote{The example of the Einstein equation is slightly different from the case of the Schrödinger equation, since the former but not the latter is best conceived as a kind of constraint equation rather than a typical dynamical law, i.e.~as time evolution. Nevertheless, we take both notions to fit under our broad umbrella term `dynamics'. For a discussion, see \citeauthor{deharobutterfieldOUP} (2025:~pp.~62--64).}
Furthermore, the state-space and the set of quantities are both usually equipped with additional structures (e.g.~symmetries) and rules for evaluating the values of quantities on physical states. These triples $\bra\mathcal{S},\mathcal{Q},\mathcal{D}\ket$ encode the basic structure of physical models, and will serve us in articulating the Schema.\footnote{Note that not all theories are thus presented: in Section \ref{KWd}, we will discuss the Kramers-Wannier duality of the Ising model, which is a probabilistic theory presented in terms of a set of states, a Hamiltonian and other quantities, and the Boltzmann weights or probabilities of states. As we will see, this is only a matter of a different  formulation of the models, rather than a limitation of the Schema below. Indeed, the Schema can also be used for these other formulations.} \\
\\
{\bf The conception of duality.} Given a theory and a pair of models thus formulated as triples,
we define a {\bf duality} as a bijective, structure-preserving, map (i.e.~an \textbf{isomorphism}),\footnote{In contrast to mathematics and mathematical physics, in model theory and logic, isomorphism is considered a very strong condition, because putatively isomorphic models, once written in a logical language, often turn out to be neither isomorphic nor logically equivalent. Their signatures are different, while isomorphism requires that the signatures are the same. Thus \citeauthor{deharobutterfieldOUP} (2025:~pp.~388--393) show that SO(2) and U(1), written in a formal language, are not in this sense isomorphic; however, their definitional extensions {\it are} logically equivalent. Likewise, \citeauthor{barrett2016glymour} (2016:~pp.~469-470) give an example of two formulations of the theory of groups that are definitionally equivalent, but are not logically equivalent. Therefore, the isomorphism criterion has sometimes been characterized as being `too strict'. \citeauthor{deharobutterfieldOUP} (2025:~Chapter 11) discuss dualities in connection with equivalence in logic, and address some, but not all, of these criticisms. While an in-depth discussion of these issues is beyond the scope of this Element, in cases where the isomorphism is not obvious because the dual models are written in different variables, one of the roles of the common core theory that we will discuss below is to provide the change of variables that enables the proof of the isomorphism. Although the details differ, this is similar to the role of sophistication as envisaged by \cite{dewar2019sophistication}, which we will discuss at the end of Chapter \ref{3}.} 
$d$, between the models' sets of states $\mathcal{S}$, and between the models' sets of quantities $\mathcal{Q}$.\footnote{To have a duality, it is sufficient that the isomorphism holds between the dynamical states. However, it is important to note that, in many examples of classical theories for which the duality map is initially found for states that satisfy each of the models' equations of motion, the map can be extended to also relate the Lagrangians or actions of the two models, where the variables do {\it not} necessarily satisfy the equations of motion, and that the Schema is well-equipped to describe these cases. For a discussion, see \citeauthor{deharobutterfieldOUP} (2025:~pp.~30--40).  This extension of the duality map to the ``off-shell'' actions is often achieved by the common core theory. For quantum theories, we can almost always think of a duality as an isomorphism of algebras and their associated sets of states and dynamics. In practice, this is usually expressed in terms of complete sets of correlation functions.} 
We define the structure to be preserved as follows: there is an isomorphism of state-spaces and an isomorphism of sets of quantities,\footnote{Note that, to have a duality, we require isomorphisms of state-spaces and sets of quantities, usually algebras. Thus the structures with which these state-spaces and algebras are equipped are also preserved.} 
which are (i) equivariant with respect to the dynamics $\mathcal{D}$; and (ii) preserve the values of the quantities. We can visualize these relations through the commutative diagram in Figure \ref{obv2}.\footnote{As one can see from the diagram, a duality is a {\it pair of maps}, $d_{\cal S}$ and $d_{\cal Q}$: one for the states, and one for the quantities. However, to simplify our discussion, we will not often need to use this notation.}
\begin{figure}
\begin{center}
\bea
\begin{array}{ccc}{\cal S}_1&\xrightarrow{\makebox[.6cm]{$\sm{$d_{\cal S}$}$}}&{\cal S}_2\\
~~\Big\downarrow {\sm{$D_1$}}&&~~\Big\downarrow {\sm{$D_2$}}\\
{\cal S}_1&\xrightarrow{\makebox[.6cm]{\sm{$d_{\cal S}$}}}&{\cal S}_2
\end{array}~~~~~~~~~~~~
\begin{array}{ccc}{\cal Q}_1&\xrightarrow{\makebox[.6cm]{$\sm{$d_{\cal Q}$}$}}&{\cal Q}_2\\
~~\Big\downarrow {\sm{$D_1$}}&&~~\Big\downarrow {\sm{$D_2$}}\\
{\cal Q}_1&\xrightarrow{\makebox[.6cm]{\sm{$d_{\cal Q}$}}}&{\cal Q}_2
\end{array}\nonumber
\eea
\caption{\small Equivariance of duality and dynamics, for states and quantities.}
\label{obv2}
\end{center}
\end{figure}

The isomorphism condition for duality secures that states and quantities are in a precise correspondence across the duality map, while the equivariance condition for the dynamics secures that the dynamics of the two models are compatible. Value-preservation for quantities is also required. (This is usually interpreted as the two theories having, for all possible measurement outcomes related by duality, the same values.) \\
\\
{\bf The common core theory.} We have defined a duality as an isomorphism between models. We now return to our earlier discussion of theories and models, where duals are, in our jargon, models. This raises the natural question of what {\it theory} these duals are models of. We shall call this theory a \textbf{common core theory} for the dual models, i.e.~one that contains the structure that is common to the duals. Thus we require that the duals are instantiations, usually mathematical representations, of this common core theory. This formal condition secures that the structure that is common to the models is  capable of physical interpretation, and so is physically significant. That it is a `theory' (albeit uninterpreted at this stage) means that it is again a triple $\bra\mathcal{S},\mathcal{Q},\mathcal{D}\ket$.\footnote{Although it is, in practice, a convenient choice that the common core theory is of the same type as the models, i.e.~a triple, this is not necessary. For example, a common core theory could be defined syntactically, using a set of axioms, rather than set-theoretic structures.} 

Since at this formal stage this is an uninterpreted theory, we also call it the {\bf bare theory}. The requirement that the common core is a theory means that being a duality is a strong requirement, because the common structure should amount to a triple of set of states, quantities, and dynamics. This implies that not any old partial isomorphism between models is a duality. Furthermore, the role of the common core theory is also that, especially in cases where the two duals are very different, it makes the isomoprhism explicit. This theory deals with structure that is invariant under the duality map $d$, similarly to how we would define a theory that is invariant under the action of a certain symmetry relating two models of that theory.\footnote{Our phrase `deals with structure that is invariant under the duality map' should not be taken to mean that the common core theory always has less structure than its models (just as the presentation of a group does not always use less structure than each of the representations of this group). For, as with symmetries, invariance can be achieved in different ways, e.g.~by `reduction' or by `sophistication' (see \citet{dewar2019sophistication} and \citet{martens2021sophistry}). Indeed, the definition of the common core theory is, in itself, entirely independent of the reduction vs.~sophistication debate, and the common core theory does not in general privilege reduction: sophistication is also possible. Thus in many cases, the variables used by the common core reflect the specific structure of the models. This corresponds to \cite{deharobutterfieldOUP}'s contrast between `abstract' and `augmented' common core theories. Thus the main point is that the common core theory does not privilege one model, with its specific structure, at the expense of another, but exhibits how the duals are isomorphic. Indeed, a major role of the common core theory is its enabling a proof of equivalence of models that is entirely obvious.} 
We then recover the original duals by representing the common core theory using \textit{specific structure}, i.e.~further, non-duality invariant bits of formalism that are used to individuate and define the models. Using an analogy from \citeauthor{deharobutterfieldOUP} (2025), we can think of the relation between dual models and their common core, in terms of the relation that obtains between for example a group and its representations. Here, the common core is like the abstract group, and each dual model is like a representation of the group: thus there is a duality if two such representations are isomorphic.\\
\\
{\bf Interpreting duals.} We have so far discussed models and theories as formal i.e.~uninterpreted triples. However, as we mentioned in our discussion of empirical and theoretical equivalence, to understand the roles of dualities in physical theories we cannot ignore questions about the relation between the duals and the systems they model, i.e.~about their \textbf{interpretation}. For our purposes, it is useful, and also a widespread philosophical practice, to think of an interpretation as a mapping of a theory or model (here, a triple $\bra\mathcal{S},\mathcal{Q},\mathcal{D}\ket$) into a domain of application $D$ in the world. The domain of application is the relevant collection of physical systems that our theory or model describes, identified through observational and experimental procedures, specific limits etc., as they are familiar from the practice of science. In other words, the domain of application of a theory or model is a part or domain of the world, of physical reality, that is suitable to be described by a given triple $\bra\mathcal{S},\mathcal{Q},\mathcal{D}\ket$.

To sum up: an interpretation is a map $i$ from a triple $\bra\mathcal{S},\mathcal{Q},\mathcal{D}\ket$ to a domain of application $D$. 
Since we are interested in studying how interpretations interact with dualities, we should ask what sorts of maps are allowed for dual models of the type described by the Schema. Based on the structure that these maps map, there are two possible kinds of interpretation:\footnote{For a related, but different, discussion of the metaphysical issues that arise when interpreting dualities, see \citeauthor{le2018duality} (2018).}
\begin{itemize}
\item \textbf{Internal interpretations:} These are interpretations that only interpret the structure that is common to the duals, i.e.~the common core theory. In other words, the interpretation map $i$ maps duals into a single domain of application $D$, so that $D$'s features depend on the structure of the common core only, and not on further structure typical of either dual, i.e.~their specific structures. Since internal interpretations only deal with structure that is invariant under the duality map, they give interpretations of the {\it common core theory}.
\item \textbf{External interpretations:} These are interpretations that, besides mapping the structure that is common to the duals (i.e.~the common core theory), also map the specific structure into the domain of application $D$. This entails that the specific structure is part of the physical content of our theories, i.e.~it is part of the structure that represents the theory’s domain of application. As a consequence, since dual models do not have this specific structure in common, any two models that are interpreted according to their external interpretations are mapped into different domains of application.
\end{itemize} 

The external vs.~internal contrast exhausts the options for interpreting duals. For either we interpret each dual with its specific structure, independent of the other dual's non-matching specific structure 
(as in an external interpretation), or we do not interpret the specific structure, and we interpret only the common core (as in an internal interpretation). Thus going back to the Schema described above, there is no further structure that we could interpret: and so, for duals and interpretations thus defined according to the Schema, we have exhausted the different ways to interpret dualities. 

We have so far given a general introduction to dualities, without delving into specific examples. In the following Chapters, we will give several examples that illustrate these ideas, and which show the variety of interesting philosophical issues that dualities raise. In the next Section, we first introduce a number of interesting philosophical and conceptual themes that emerge from the study of dualities, and which illustrate dualities' many roles in physics: these themes will guide our analysis of various examples in the rest of this book.

\subsection{Dualities in Physics: Themes, Roles and Types}\label{2.2}

The previous Section introduced the Schema as a formal way to think about dualities, together with its natural interpretative options. In this Section, we go on to discuss the main themes associated with dualities, roles that they play in both physics and philosophy, and main types of dualities. 

We begin by discussing the {\bf themes}: we will emphasize three main ones that will recur in the rest of this book, and which our examples will illustrate. These themes are contrasts that characterise dualities, and we will label them as follows: \textit{hard-easy}, \textit{elementary-composite} and \textit{exact-effective}.\\
\\
{\bf Hard-easy.} The contrast `hard-easy' means that a problem that is difficult to solve in one dual, is easy to solve in the other dual. For example, a difficult calculation becomes tractable after a duality transformation. This is similar to how changes of variables can be valuable tools in solving integrals or differential equations.

To see this more precisely, consider how dualities map the coupling constants of physical theories.\footnote{An example of this inversion of coupling constants, in Section \ref{EMMa}, is \textit{electric-magnetic} duality. As we will discuss in Section \ref{3.1} in the example of position-momentum duality in quantum mechanics, the {\it hard-easy} theme is not always related to the inversion of a coupling constant. Indeed, this latter way of illustrating the theme is typical of more advanced examples in quantum field theory and string theory.} 
Recall that coupling constants characterise the strength of interactions: the larger the value of the theory’s coupling constant, the stronger the interactions. %A simple example is Newton’s constant, $G_{\tn N}$.
%\bea\label{eq:newteq}
%F = G_{\tn N}\,\frac{m_1m_2}{r^2}\,.
%\eea
%Newton’s constant here contributes to the strength of the gravitational force $F$ by making the right-hand side of this equation larger or smaller according as Newton's constant is larger or smaller, so that the gravitational force $F$ is accordingly stronger or weaker.
%Since coupling constants govern the strength of the interactions, 
Thus we will call a (regime of a) theory where interactions are weak, \textit{weakly coupled}. Likewise, a (regime of a) theory where interactions are strong, is called \textit{strongly coupled}. Furthermore, if we take the coupling constant of a theory to be an arbitrary positive real number then, according to our usage of `theory' and `model', each of these regimes gives a model, i.e.~a specific realization of the theory. Each such model is identified by its specific structure: here, a choice of coupling.\footnote{There is here a judgement call about how general a theory should be taken to be: should its coupling parameter be an arbitrary positive number, or should it be fixed to a specific value? Physicists usually follow the former convention, while in philosophy such parameters are often taken to be fixed. We here follow physicists' conventions. Although nothing of substance hinges on this at this level, it is a useful choice when discussing dualities.}\label{convention}

While the weakly-coupled regime of a theory usually allows perturbative calculations, the strongly-coupled regime rarely allows this, and often remains poorly understood. 

%In quantum field theory, this greater control over the weakly-coupled regime stems from the fact that perturbative methods to approximate a solution fail at strong coupling, and so are only applicable in the weakly-coupled regime. 
Perturbation theory is a method to approximate the value of physical quantities for an (unknown) solution, by taking their values in another (known) solution and adding small perturbations. `Small' is here defined by a parameter such as a coupling constant. This amounts to expanding the value of the quantity, for example a scattering amplitude $A(g)$ with coupling constant $g$, as a power series in $g$ around the value at zero coupling, i.e.~$A(g=0)$:
\bea\label{eq:pert}
 A(g) = \sum_{n=0}^{\infty} a_n\, g^n.
\eea
Since the theory can be solved exactly at zero coupling, this method gives us a way to approximate unknown solutions of quantum field theory to the ones at zero coupling. The main limitation of perturbation theory is that the series \eqref{eq:pert} is often \textit{asymptotic}, i.e.~it diverges at large values of $n$. This requires us to truncate the expansion \eqref{eq:pert} at a finite value of $n$, thus rendering the result of this procedure an approximation rather than a convergent series.

The expansion Eq.~\eqref{eq:pert} usually only works in a small neighbourhood of the known solution, $A(0)$. For large values of the coupling constant, i.e.~in the strongly-coupled regime, the expansion Eq.~\eqref{eq:pert} breaks down, and we cannot rely on perturbation theory. For this reason, the strongly-coupled regime is also called the {\bf non-perturbative} regime of the theory. Since perturbation theory is the main method to construct solutions of quantum field theories, this implies that the strongly-coupled regime is in general beyond the scope of current calculational techniques, which almost always rely on perturbation theory.

With this in hand, it is straightforward to illustrate the {\it hard-easy} theme of dualities. The point of this contrast, and the reason why dualities are important calculational and descriptive tools, is that they allow us to relate a perturbative, weakly-coupled, description in one dual, to a strongly-coupled, non-perturbative, description in the other dual. Indeed, as we will discuss in Chapters \ref{3} and \ref{DQFTG}, dualities often map the coupling constant $g$ of one dual to the inverse coupling constant, $1/g$, of the other dual. And by our discussion above, this means that a weakly-coupled calculation is dual to a strongly-coupled one. Thus using dualities we can extend the regime of applicability of our calculational techniques to non-perturbative phenomena, thus opening a new avenue for the study of the strongly-coupled regime of quantum field theories.\footnote{For an example of the use of dualities to study the non-perturbative behaviour of a quantum field theory, see \citet{vergouwen2024supersymmetry}.}\\
\\
{\bf Elementary-composite.} This is another interesting theme associated with dualities. Although we will give a more detailed treatment of this theme in Section \ref{sec: EMQFT}, we briefly sketch it here. Dualities sometimes map objects that are simple, i.e.~that cannot be decomposed into more elementary components in a given model, into objects that are composites, i.e.~that {\it can} be decomposed in the dual model. So by this theme, dualities map objects that lack mereological structure,\footnote{By mereology in philosophy, we mean the formal study of \textit{part-whole} relations.} 
into objects that have it. And this casts doubt on the fundamentality of such mereological notions, and on their ability to ``carve nature at the joints'' (see \citeauthor{castellani2017duality} (2017) for more on this point).

An especially important example of this theme is \textit{particle-soliton} duality, which maps a fundamental degree of freedom in one model to a soliton in the other model, i.e.~a topologically non-trivial configuration of the fundamental degrees of freedom of the dual. This relation is clearly an example of our theme of {\it elementary-composite}, since it relates a fundamental, elementary object, a particle, to a complex, composite one, i.e.~a soliton. A simple illustration of this phenomenon is the \textit{Sine Gordon-Massive Thirring model} duality.

Without going into the details,\footnote{See \citeauthor{de2018schema} (2018) for a detailed philosophical analysis of this case and of the closely related topic of \textit{bosonization}.} 
the point of such a duality is that we relate the fundamental fermionic degrees of freedom of the Massive Thirring model, to a topologically complex configuration of the Sine Gordon model:\footnote{This configuration is technically a kink.} 
and vice versa, we relate the fundamental bosonic degrees of freedom of the Sine Gordon model to a composite of two fermion fields in the Massive Thirring model. Due to this mapping between bosons and fermions, this duality is an example of \textit{bosonization}. The relation between the Sine Gordon and Massive Thirring models illustrates the theme of {\it elementary-composite}, since it relates simple fundamental entities to composite ones, and vice versa, thus raising important questions about the meaningfulness of such a distinction.\\
\\
{\bf Exact-effective.} In some cases, what looks like a duality may in fact not be a precise isomorphism between models. For example, the map may only map some quantities from one model to the other, but not all of them, as is required for a duality. Also, while a duality is by definition \textit{exact}, i.e.~valid without approximations, there is an {\bf effective duality} when the map between models approximates a duality in a regime of parameters. More generally, a {\bf quasi-duality} is any relation between models that is similar to or close to a duality, but is not a duality, i.e.~a precise isomorphism of the type defined in Section \ref{2.1}. Most examples of this theme are somewhat technical, and so we will not here give an intuitive introduction like we did for the other themes (important examples of this are in Section \ref{sec: EMQFT}: particle-soliton duality, dual superconductivity, and electric-magnetic duality for ${\cal N}=2$ supersymmetric Yang-Mills theory). However, to get an idea of the type of physics involved, note that arguably the most important and famous instance of effective duality is the relation between perturbative string theory and non-perturbative M-theory, i.e.~the fundamental theory underlying the five different superstring theories, that we will discuss in Section \ref{sec: stringduality}.

An immediate consequence of the fact that quasi-dualities are not isomorphisms of models, so that there is no precise common structure of models that they should preserve, is that there is no straightforward common core theory underlying the two quasi-dual models. (We say `no straightforward common core theory', since a precise common core theory for duals might exist in some limit of parameters in which we do recover a duality.) 

The main reason for the interest in quasi-dualities, which we will explore in Section \ref{QD}, is that they suggest a novel way to represent the structure of scientific theories. Indeed, quasi-dualities suggest that scientific theories are not best represented as e.g.~collections or sets of models (as in the semantic conception of theories), but rather as {\it differential manifolds} or more general geometric objects (such as algebraic varieties). We will call this view of physical theories the {\bf geometric view} of theories.\\
\\
We next discuss some of the {\bf roles} that dualities play in physical theorizing. Here the main distinction, introduced in \citeauthor{de2019heuristic} (2019), is between the \textit{theoretical} and \textit{heuristic} roles of dualities. In simple terms, the relevant distinction is between the role of dualities as well-established theoretical relations that can be used to study the properties of scientific theories (their theoretical role), and dualities as clues for the construction of more fundamental theories (their heuristic role).

It is worth mentioning two central philosophical topics to which dualities, in their {\bf theoretical role}, are closely related: (i) {\it theoretical equivalence}, which will be the topic of  Section \ref{TE}; and (ii) {\it scientific realism}, in particular in connection with issues of under-determination. While we will not discuss topic (ii) in detail in this book, we refer the reader to \citeauthor{matsubara2013realism} (2013), \citeauthor{read2016interpretation} (2016), \citet{le2018duality}, \citet{de2023empirical} and \citet{deharobutterfieldOUP} for thorough treatments. 

About the {\bf heuristic role}, it is worth mentioning that the natural heuristic use of dualities is as guesses for the minimal structure of a more fundamental theory, of which duals are expected to be some kind of approximation or limit: we call this deeper theory a \textit{successor theory}. If the duals are approximations or limits, then it is natural, from the point of view of the more fundamental theory that we are constructing,  to think of the duality as being a limit of an effective duality or a quasi-duality. Thus in their heuristic role, dualities can often be taken to point towards a more fundamental theory. 

Finally, we discuss the special {\bf types} of dualities that one finds in the literature, and which we will discuss in this book. There are two special types: \textit{quantum} dualities and \textit{self}-dualities. %These are in addition to standard dualities, which for us means any isomorphism between models represented as triples and as in the Schema described in Section \ref{2.1}.

A {\it quantum duality} is a duality between quantum mechanical models whose classical limits are {\it not duals}. These dual quantum models are different representations of a single quantum theory, and so the two classical models can often be seen as different classical limits of a single quantum theory.
%Thus different-looking classical models are classical limits of a single quantum. The two duals, as {\it classical} i.e.~non-quantum models, are not duals at all. Thus quantum dualities imply the existence of a single quantum description with at least two classical limits.

A {\it self-duality} maps a given model onto another model of the same kind, or onto the same model, while it also changes a parameter (possibly with other additional changes that compensate for this variation). An example of this is the self-duality of $\mathcal{N}=4$ supersymmetric Yang Mills theory that we will discuss in Section \ref{sec: EMQFT}, where the relevant parameter is the coupling $g$, and the additional changes are the exchange of the theory's gauge group $G$ with its Langlands dual group $G^{\lor}$ (but this last change will not play any role in our discussion). The reason for the name `self-duality' is the physics convention of treating theories that differ only by a change in a parameter as being the same theory.\footnote{For a discussion, see footnote \ref{convention}.} 
Thus, as with symmetries, such transformations can be seen as automorphisms of the state-space and set of quatities (as we defined them above, following \citeauthor{caulton2015role} (2015)); hence the \textit{self} in `self-duality', and in this case we just talk about `the theory'.

This said about the themes, roles and types of dualities, the next two Chapters will illustrate these general ideas in several examples. Then we will return, in Chapter \ref{philiss}, to a general discussion of the philosophical questions.

\section{Classic Examples of Dualities}\label{3}

This Chapter introduces three simple examples of dualities, which will be a gentle introduction to the main features of dualities, illustrate their themes, and be a stepping stone towards the more advanced examples in the next Chapter.

Section \ref{3.1} introduces \textbf{position-momentum} duality, as a quintessential example that illustrates basic features of dualities. An important reason for choosing position-momentum duality as our first example is that it is a well-known duality, whose features will recur in the more advanced examples. 

Section \ref{KWd} then discusses \textbf{Kramers-Wannier} duality, i.e.~the duality between high- and low-temperature Ising models, which we choose for two main reasons: first, because it gives us an example of a duality involving a probabilistic theory, i.e.~the states and quantities of this theory do not satisfy a deterministic dynamics, but rather they are distributed according to a probability distribution (namely, the Boltzmann weights). Second, Kramers-Wannier duality illustrates a variety of important themes related to dualities, especially the \textit{elementary-composite} distinction.

Finally, Section \ref{EMMa} discusses \textbf{electric-magnetic} duality in the Maxwell theory. This is our first example of a duality in a deterministic field theory; and, given its structural similarities with dualities in quantum field theory and string theory, it is a natural starting point for our discussion of more advanced examples.

\subsection{Position-Momentum Duality in Quantum Mechanics}\label{3.1}

This Section discusses Fourier duality in non-relativistic quantum mechanics, and some of the philosophical questions that it raises. This example is one of the most natural entry points into dualities, since all that is required to understand it is the mathematics of Hilbert spaces and unitary maps between them. Fourier duality then amounts to the claim that there is a unitary map, the Fourier transformation, that relates the Hilbert space of a quantum mechanical system written in the position basis to the Hilbert space of the system written in the momentum basis. 

Note that, for the purposes of this discussion, we are treating the position and momentum bases as giving in principle distinct Hilbert space representations of the quantum system and its algebra of operators. Given (i) the common understanding of unitary equivalence as sufficient to establish theoretical equivalence between two Hilbert space representations, and (ii) our discussion of the position and momentum representations as apparently distinct descriptions of a system, the existence of the Fourier transformation as a unitary map amounts to the claim that Fourier duality is an isomorphism between two apparently distinct models (in the terminology of Chapter \ref{dualint}). Thus the Fourier transformation is indeed a duality, in the sense of the Schema in Section \ref{2.1}.

While Fourier duality is an exceedingly simple example, its basic structure, and in particular the idea of a duality as a unitary map between Hilbert spaces, echoes the way dualities are implemented in more advanced examples, especially in quantum field theory (see Chapter \ref{DQFTG}). However, it is worth noting another feature of Fourier duality that relates it to more sophisticated quantum field-theoretic examples: namely, the assumption that the position and momentum representations are treated as apparently distinct descriptions, especially before their unitary equivalence, in terms of a unitary map between the respective Hilbert spaces, is established. This assumption might seem artificial in the extremely simple context of non-relativistic quantum mechanics: and indeed, the existence of the celebrated Stone-von Neumann theorem would seem to confirm this impression, since it proves the unitary equivalence of all Hilbert space representations of the operator algebra of a quantum system with a finite number of degrees of freedom.\footnote{However, see \citeauthor{earman2023revealing} (2023) for subtleties regarding this claim, even in the case of a finite number of degrees of freedom.} 

However, the Stone-von Neumann theorem breaks down for an infinite number of degrees of freedom. This implies that a quantum field theory can have \textit{unitarily inequivalent} representations, and so there is nothing artificial about treating different representations as a priori distinct.\footnote{See \citeauthor{ruetsche2011interpreting} (2011) for ample discussion of the foundational and philosophical implications of this fact for our understanding of quantum field theory.} 
In this light, our treatment of Fourier duality as a duality, and of position and momentum as a priori distinct representations, renders the simple case of Fourier duality useful to understand the more advanced quantum field-theoretic examples that we will discuss later on.

With this in mind, we now examine the structure of Fourier duality in non-relativistic quantum mechanics in greater detail.\footnote{Here we follow the treatment in \citeauthor{jordan2012linear} (1969:~Sections 14-18).} 
The first element required is the notion of a state-space in quantum mechanics, namely a Hilbert space. The reason for this is straightforward: according to the Schema in Section \ref{2.1}, a duality is an isomorphism of the states, and of the quantities, of dual models. %Thus to conceptualise the Fourier duality, we first need to introduce the notion of a Hilbert space.

A Hilbert space is a complex vector space endowed with a complex inner product: this inner product can be used to define a norm, from which a topology can be defined. Thus a more precise characterization of a Hilbert space is as a \textit{topological} vector space with a complex inner product, where the norm and the topology of the Hilbert space are both induced from the inner product. A standard example of Hilbert space, which played an important role in the early days of quantum mechanics (in particular, in Schrödinger’s formulation of wave mechanics) is the Hilbert space of square-integrable functions $L^2(\mathbb{R})$, i.e.~the space of functions such that: $L^2 (\mathbb{R}) := \{ [\psi] | \psi : \mathbb{R} \rightarrow \mathbb{C}, \int_{\mathbb{R}} \dd x\, |\psi|^2 < \infty \}$ (where the equivalence relation, indicated by the square brackets, is almost everywhere equality).

Our discussion of Fourier duality requires giving two specific forms of the Hilbert space of square integrable functions: namely, a position and a momentum representation. We start with the {\it position representation}: it takes position as its fundamental variable, and defines the Hilbert space as the space of square-integrable functions of the position variable in Euclidean space $\mathbb{R}^3$, i.e.~$L^2(\mathbb{R}^3)$. Thus an element of $L^2(\mathbb{R}^3)$ is given by the \textit{wave-function} $\psi(\textbf{x})  = \psi(x_1,x_2,x_3)$, where $\textbf{x}$ is a shorthand for $(x_1,x_2,x_3)$ and labels a point in Euclidean space  $\mathbb{R}^3$. The inner product on this space is given by $\braket{\psi|\phi} = \int_{\mathbb{R}^3} \dd^3x\, \psi^* (\textbf{x}) \,\phi(\textbf{x})$. In the position representation, we define the position operator as $X_l$, with $l = 1,2,3$, as: $(X_l \,\psi)(\textbf{x}) = x_l\, \psi(\textbf{x})$; and we also define the operator that is canonically conjugate to $X_l$, i.e.~$P_m$ for $m = 1,2,3$, as: $(P_m\,\psi)(\textbf{x}) = -i\hbar\frac{\partial}{\partial x_m}\,\psi(\textbf{x})$. $P_m$ is called a `momentum operator'. While it is not necessary for our purposes to make this interpretation precise, it is useful to recall that the reason for this statement is that $X_l$ and $P_m$ obey the canonical commutation relation: $[X_l, P_m] = i\hbar\delta_{lm}$.

The {\it momentum representation} is defined by analogy with the position representation, the main difference being that we define quantum states in terms of functions over \textit{momentum space} rather than position space, i.e.~a space where each point is labelled by a vector $\textbf{p} = (p_1,p_2,p_3)$ of the momentum of the system under study, in each spatial direction. With this Hilbert space at hand, we can then define a momentum operator $P_m$ as follows: $(P_m\, \psi)(\textbf{p}) = p_m \,\psi(\textbf{p})$. Furthermore, we can also identify the operator canonically conjugate to $P_m$ as $X_l$: $(X_l\,\psi)(\textbf{p}) = i\hbar\frac{\partial}{\partial p_l}\,\psi(\textbf{p})$. Analogously to the position representation, we can think of $X_l$ as a position operator, since it stands in the canonical commutation relation to the momentum operator $P_m$.

We are here interested in the Fourier transformation, $\mathcal{F}$, as a map that relates the position and momentum representations. This map is given as follows: 
\bea\label{eq: fourier1}
(\mathcal{F}\psi)(\textbf{p}) = (2\pi\hbar)^{-\frac{3}{2}} \int_{\mathbb{R}^3} \dd^3x\, e^{-\frac{i}{\hbar}\textbf{p}\cdot\textbf{x}}\, \psi(\textbf{x})
\eea
\bea\label{eq: fourier2}
\psi(\textbf{x}) = (2\pi\hbar)^{-\frac{3}{2}} \int_{\mathbb{R}^3} \dd^3p \,e^{\frac{i}{\hbar}\textbf{p}\cdot\textbf{x}}\, (\mathcal{F}\psi)(\textbf{p})\,.
\eea
Note that the Fourier transformation $\mathcal{F}$ in Eq.~\eqref{eq: fourier1} expresses the momentum representation data in terms of the position representation, while its inverse $\mathcal{F}^{-1}$ in Eq.~\eqref{eq: fourier2} expresses the data of the position representation in terms of the momentum representation. Furthermore, by \textbf{Plancherel’s theorem}, $\mathcal{F}$ preserves inner products:
\begin{itemize}
\item[] \textbf{Plancherel’s theorem}: $\mathcal{F}$ preserves norms and has an inverse. Hence, $\mathcal{F}$ is a \textit{unitary} map. As a consequence, $\mathcal{F}$ preserves inner products, i.e.~$\braket{\mathcal{F}\psi|\mathcal{F}\phi} = \braket{\psi|\phi}$.
\end{itemize}

\noindent{\bf Fourier duality and the Schema.} We now discuss how Fourier duality counts as a duality according to the Schema. First, take the Hilbert spaces in the position and momentum variables, discussed above, to be our dual models (in the sense of `model' in Chapter \ref{dualint}). Second, take the duality map between these two models to be the Fourier transformation, i.e.~$d_{\mathcal{S}}=\mathcal{F}$. Then: (i) In virtue of $\mathcal{F}$'s being a unitary equivalence of Hilbert spaces, it is an isomorphism of the state-spaces $\mathcal{S}$. (ii) $\mathcal{F}$ also induces an isomorphism between the sets of operators, i.e.~the sets of quantities $\mathcal{Q}$. For example, the Hamiltonian in the momentum representation and the Hamiltonian in the position representation are related as follows: $H_{\tn{momentum}} = \mathcal{F}H_{\tn{position}}\,\mathcal{F}^{-1}$. (iii) $\mathcal{F}$ is also equivariant with respect to the dynamics of the system. This can be seen from the relation between the Hamiltonian in the position and momentum representation given under (ii), \footnote{Indeed, insofar as the Hamiltonian is just another operator in the quantum theory, this dynamical equivalence is a special case of the equivalence between the quantities $\mathcal{Q}$ in the two representations.} 
which also maps the Schr\"odinger equation of one model to the Schr\"odinger equation of the other model. Therefore, this establishes the dynamical equivalence of the models. (iv) Finally, note that the preservation of inner products, secured by the unitarity of $\mathcal{F}$, guarantees that the values assigned by the states $\mathcal{S}$ to each quantity $\mathcal{Q}$ are the same in both representations.

Having discussed how Fourier duality illustrates the Schema, it is interesting to discuss in more detail the {\it common core theory} underlying the two Fourier dual models. The answer to this question has been known since von Neumann's seminal book (\citeauthor{von1955mathematical}, 1932): namely, quantum mechanics as described in terms of Hilbert spaces. For once we start representing quantum mechanics in terms of Hilbert spaces, it is natural to associate a specific quantum system with a Hilbert space of states for that system, which we will call $\mathcal{H}$. The position and momentum representations are then realized as two different bases of the Hilbert space, and the Fourier duality between them is a change of basis. The Hilbert space is itself independent of the choice of a basis, so that it can be taken to be the state-space $\mathcal{S}$ of the common core theory, of which the two basis-dependent state-spaces (whose states are written in Eqs.~\eq{eq: fourier1} and \eq{eq: fourier2}) are representations. Likewise, the algebra of operators\footnote{More precisely, the algebra of non-relativistic quantum mechanics, of which the self-adjoint elements give the theory’s observables.} 
associated with that Hilbert space can be taken to be the set of quantities, $\mathcal{Q}$, of the common core theory, of which the position and momentum operators, written above in terms of a choice of basis, are representations. Finally, the dynamics $\mathcal{D}$ is given by the Hamiltonian (i.e.~the time-evolution) operator on the Hilbert space. Thus the common core theory for Fourier duality is Hilbert space quantum mechanics.

Fourier duality also illustrates our theme of {\it hard-easy} from Section \ref{2.2}: namely, that some problems that are hard to solve in one representation, are easy (or at least easier) to solve in the other. Depending on the form of the Hamiltonian, this is the case when the Schr\"odinger equation is easier to solve in one representation than in the other.
For example, for constant potentials (such as the one-dimensional infinite square well, or the step potential) the Schr\"odinger equation is easier to solve in the position representation. For the time-independent Schr\"odinger equation is then the free wave equation, and the boundary conditions have a clearer interpretation in position space. The momentum representation has advantages in other cases: for example, in three-dimensional problems with symmetries, like the rigid rotor, it is easier to think about the wave-functions as eigenstates of the angular momentum operators. And for scattering problems on lattices, it is also often easier to work with momentum vectors on the reciprocal lattice and then Fourier transform to the original lattice.

\subsection{Kramers-Wannier Duality}\label{KWd}

The Ising model on a two-dimensional homogeneous square lattice is defined by its Boltzmann weights, $e^{-\beta H}$, where the Hamiltonian is given by the pairwise interactions between the spins located on the vertices of the lattice:
\bea
H=-J\sum_{ij}s_is_j\,.
\eea
The spins $s_i$ can take the values $\pm1$, and the little roman labels $i,j$ label the vertices of the lattice (also called `lattice sites'). The sum is over pairs of nearest-neighbour sites on the lattice, i.e.~neighbouring spins. $J$ is a constant that is interpreted as the coupling strength between nearest neighbours. 

We will here focus on the case $J>0$: in this case, the energy is lowest for neighbouring spins that have the same sign, and so (if the temperature is not too high) the tendency for nearest neighbours is to be aligned, i.e.~to point in the same direction. And since the spins are aligned, the lattice is {\it ordered}, which manifests itself in the non-zero value of the {\it spontaneous magnetisation} $M$, which is the average value of the spins on the lattice. Such a lattice gives a good model of a {\bf ferromagnet}, with spins over large (macroscopic) regions pointing in the same direction, and with a non-zero spontaneous magentisation that couples to external electric and magnetic charges and currents in the familiar way.

If an external magnetic field were applied to the ferromagnet, the spins would tend to align in the direction of the magnetic field, increasing the net magnetisation. Thus the word `spontaneous' indicates that the magnetisation of a ferromagnet can be non-zero even in the absence of an external magnetic field.

Note the important qualification `at low temperatures'. For if the temperature increases, the individual contributions of neighbouring pairs to the energy become less important, and larger regions will appear where the spins are disordered, i.e.~not correlated, with values that are randomly distributed. High temperature tends to wash out the ordering of the spins. The spontaneous magnetisation is then zero or close to zero. This is called a {\it paramagnetic phase} of the ferromagnet, where the magnetisation is lost due to the high temperature.

For real magnets, the two phases are not neatly separated, and there are also other effects, such as {\it hysteresis}, where the magnetisation is a multi-valued function of the temperature and the external magnetic field. The material ``retains a memory'' of the external magnetic field that was applied.

But in the thermodynamic limit of the simple Ising model, the two phases {\it are} neatly separated. At temperatures lower than the critical temperature $T_{\sm c}$, the spontaneous magnetisation is non-zero, the spins are ordered, i.e.~aligned along a preferred direction, and so the material does not have a macroscopic symmetry (this is called the `broken symmetry' phase). As we approach the critical temperature, the spontaneous magnetisation goes smoothly to zero, and it remains zero above the critical temperature. The spins are disordered, and, since they are not aligned along a preferred direction, the material has a macroscopic symmetry. 

The duality says that the partition function of the Ising model at the inverse temperature $\b:=J/k_{\tn B}T$, and the partition function at the {\it dual} inverse temperature, $\ti\b:=\ti J/k_{\tn B}\ti T$, are related as follows:
\bea\label{KWdual}
%\frac{Z(\b)}{\sinh^{N/2}2\b}=\frac{Z(\ti\b)}{\sinh^{N/2}2\ti\b}\,,\,\,\,\,\,\,\,\,\,\,\,\,\,\,\sinh2\b\sinh2\ti\b=1\,,
\frac{Z(\b)}{\sinh^{N/2}2\b}=\frac{Z(\ti\b)}{\sinh^{N/2}2\ti\b}\,\\
\sinh2\b\sinh2\ti\b=1\,,\label{bbeta}
\eea
where $N$ is the number of spins on the lattice. The second line relates the inverse temperature to its dual: and it follows from this relation that, when $T$ is high, the dual temperature $\ti T$ is low, and vice versa. 

The above duality relation is surprising, because it relates the value of the partition function of the Ising model to that of its dual (i.e.~defined at the dual inverse temperature) or, alternatively, low- and high-temperature Ising models. Thus, despite the physical differences between the two phases, their partition functions have the same form and take the same numerical values for dual values of the temperatures, i.e.~values related by Eq.~\eq{bbeta}.

The duality relation, Eq.~\eq{KWdual}, can be generalised to other statistical quantities defined from the partition function. For example, just like the Ising model has a magnetisation $M$ that is a function of the temperature, the dual Ising model has a dual magnetisation, $\ti M$. As we discussed above, in the thermodynamic limit, the magnetisation is non-zero for $T<T_{\sm c}$, and zero for $T>T_{\sm c}$. For this reason, the spontaneous magnetisation is called an {\bf order parameter}, since its non-zero value indicates the ordered, broken symmetry, phase, and its zero value indicates the disordered, symmetric, phase. This is opposite for the dual magnetisation, which is non-zero in the disordered, symmetric, phase: and zero in the ordered, broken symmetry, phase. Thus the dual magnetisation is called a {\bf disorder parameter}.

Kramers and Wannier used the duality relation, Eq.~\eq{KWdual}, to find the value of the critical temperature $T_{\sm c}$. This is the value at which the free energy has a (single) singularity, which happens when the inverse temperature and its dual have the same value, i.e.~for $\b=\ti\b$.

This ability to relate, at least in some salient cases and in an approximate sense, different macroscopic phases of systems (here, an ordered and a disordered phase) by using a duality, is one of the heuristic virtues of dualities, and illustrates their {\it heuristic role} (see Section \ref{2.2}). For it can be difficult to describe the disordered phase of a system using the model normally associated with that system: for example, because the order parameter is zero when the temperature is high or the coupling is strong, so that it gives no detailed information about the behaviour of the system in the disordered phase. Then it is useful to go to a dual description where the system has a dual temperature or a dual coupling that is low, so that e.g.~perturbative methods can again be applied. In such a dual description, a dual order parameter may give detailed information about the disordered phase. 

Since the two duals of Kramers-Wannier duality are both Ising models, but written in different variables, the common core theory of Kramers-Wannier duality is the Ising model itself. Indeed, this example differs from e.g.~position-momentum duality and electric-magnetic duality in that the two duals already have the same form, and so they are obviously isomorphic: namely, they are both Ising models. Thus  in this case there is no practical gain in introducing a common core.\footnote{Having said that, we note that, although this common core theory has a physical interpretation as an Ising model, this interpretation cannot commute with the duality map, because one dual is at high temperature while the other dual is at low temperature. This means that our `internal' interpretation does not assign a definite value to the temperature. Note that this is consistent with the fact that the common core is not a bare theory, and it only implies that there is no interpretation of this common core as a physical Ising model in a way that is compatible with the duality.}

\subsection{Electric-Magnetic Duality in the Maxwell Theory}\label{EMMa}

Our third and last example is the electric-magnetic duality of the Maxwell theory: this is a duality for a classical deterministic field theory, and its scientific importance lies in its being an exemplar for dualities in quantum field theory and string theory. This is witnessed by the recurrence of electric-magnetic duality in the discussions in Chapter \ref{DQFTG}.

The simplest case of electric-magnetic duality is the invariance of the set of four Maxwell equations in vacuum:\footnote{More detailed introductions are in \citet{olive1997introduction} and \citet{figueroa1998electromagnetic}. For a philosophical discussion of the interpretation of electric-magnetic duality, see
\citet{weatherall2020equivalence}.} 
\bea\label{4Max}
\nabla\cdot{\bf E}=0\,,&&\na\times{\bf E}+~~\frac{\pa{\bf B}}{\pa t}~=0\nn
\nabla\cdot{\bf B}=0\,,&&\na\times{\bf B}-\frac{1}{c^2}\frac{\pa{\bf E}}{\pa t}=0\,.
\eea
These equations are invariant under the following exchange of the electric and magnetic fields:
\bea\label{EMd}
{\bf E}/c&\mapsto&{\bf B}\nn
{\bf B}&\mapsto&-{\bf E}/c\,.
\eea
This is the simplest statement of {\bf electric-magnetic duality} for the Maxwell theory. The {\it state-space} $\mathcal{S}$ is the space of electric and magnetic field configurations, respectively ${\bf E}$ and ${\bf B}$ (see Section \ref{2.1}). This state-space is mapped onto a dual space of states, $\mathcal{S}'$, whose electric field is ${\bf E}'={\bf B}c$, and the magnetic field is ${\bf B}'=-{\bf E}/c$. 

This exchange also maps the {\it quantities} of the two models. For example, the energy, the Poynting vector, and the stress-energy tensor of the two models are mapped into each other, and they take the {\it same values} on the corresponding states. 

It will be useful to reformulate the above electric-magnetic duality in manifestly Lorentz invariant form, because this will cast light on more advanced dualities in the next Chapter.

Thus we introduce a gauge field $A$ (i.e.~a one-form potential) and its corresponding Faraday tensor, i.e.~the curvature two-form $F=\dd A$. The four Maxwell equations are then summarised as follows:
\bea\label{Biaeom}
\mbox{Bianchi:}\,\,\,\,\,\,\,\,\,\,\,\,\dd F&=&0\nn
\mbox{Eom:}\,\,\,\,\,\,\,\dd*F&=&0\,,
\eea
where $*$ is the Hodge star that defines the {\it Hodge dual} of the Faraday tensor. In components, the Hodge dual is given by:
\bea
(*F)^{\m\n}=\half\e^{\m\n\l\s}F_{\l\s}\,,
\eea
where $\e$ is the completely antisymmetric epsilon-tensor in four dimensions. The first line in Eq.~\eq{Biaeom} is the Bianchi identity that follows from the fact that the Faraday tensor is an exact two-form, i.e.~$F=\dd A$. The second line is the equation of motion, which can be derived by varying the Maxwell action. Together, these two tensorial equations summarise the four Maxwell equations.

{\bf Hodge duality} then exchanges the Faraday tensor with its Hodge dual, as follows:
\bea
F\mapsto*F\,.
\eea
As one can check by writing out the components of the Faraday tensor in a Lorentzian system of coordinates, the Hodge duality formula is the Lorentz-covariant version of the electric-magnetic duality formula for the corresponding tensor components, i.e.~Eq.~\eq{EMd}. 

The Hodge duality symbol satisfies $*^2=-1$.\footnote{To check this fact, one uses the formula for the contraction of two totally antisymmetric $\e$-tensors, together with the fact that the signature is Lorentzian.} 
Using this property, one can verify that Hodge duality exchanges the Bianchi identity and the equation of motion in Eq.~\eq{Biaeom} and vice versa, so that this set of two equations is invariant. 

So far we have discussed the Maxwell equations in vacuum. If we wish to introduce charges into the theory, we require both electric and magnetic charges so that one will be mapped into the other by the duality, and electric-magnetic duality will be respected. 

Dirac famously introduced magnetic monopoles into electrodynamics. He argued that, in the presence of monopoles, the single-valuedness of the wave-function requires that the electric charge $e$ and the magnetic charge $g$ are related in the following way:\footnote{See \citet{dirac1931quantised} and \citet{dirac1948theory}.}
\bea\label{Diracq}
eg=2\pi n\hbar\,,
\eea
where $n$ is a (positive or negative) integer. This is the {\bf Dirac quantisation condition}, and there are two things to note about it: first, that the electric and magnetic charges are each other's reciprocals, so that if $e$ is small relative to a relevant measure of electric charge, then the magnetic charge $g=2\pi n\hbar/e$ is large, and vice versa. (This is consistent with the fact that magnetic monopoles have not been observed in nature: the attractive magnetic force between monopoles of opposite charge is much larger than the Coulomb force of two electrons, and so monopoles tend to associate in pairs.) 

Second, the Dirac quantisation condition, Eq.~\eq{Diracq}, is invariant under electric-magnetic duality, which acts on the charges as: $e\mapsto g$ and $g\mapsto-e$ (and $n$ also changes sign).\footnote{For simplicity, and since it does not affect our argument, we temporarily use units where $c=1$.} 
Thus, because of the inversion of the charge and the appearance of $\hbar$, electric-magnetic duality is an example of a {\it weak-strong coupling duality}, and so it illustrates the {\it hard-easy} theme from Section \ref{2.2}. As we will discuss in the next Chapter, other versions of electric-magnetic duality also illustrate the {\it elementary-composite} theme.\\
\\
{\bf The common core theory.} For the simple Maxwell theory, perhaps the simplest way to formulate a common core theory is by defining the following complex vector field:
\bea
\mathcal{E}:={\bf E}/c+i{\bf B}\,.
\eea
In terms of this complex vector, the Maxwell equations in vacuum, Eqs.~\eq{4Max}, reduce to the condition that $\mathcal{E}$ is divergence-free, and satisfies the following linear wave equation:
\bea
\nabla\times\mathcal{E}={i\over c}\,{\partial \mathcal{E}\over\partial t}\,,
\eea
which in turn implies that $\mathcal{E}$ satisfies the Klein-Gordon equation. 

The duality map rotates this vector field over $\pi/2$ in the complex plane: $d_{\cal S}:\mathcal{E}\mapsto -i\mathcal{E}$.\footnote{In fact, any U(1) transformation, i.e.~any rotation in the complex plane, leaves the Maxwell equations invariant, and maps states and quantities in the correct way, so that it is a duality transformation.}
This means that, formally, we can think of the duality transformation as a rotation of the vector in the complex plane: or, alternatively, we can adopt a passive interpretation where the duality transformation corresponds to a choice of coordinates (specifically, a choice of complex structure on the complex plane). In what follows, we will adopt this latter interpretation.

The states of this theory are states of a complex, transverse-polarised, vector field whose four-momenta are null vectors (so that each plane wave propagates at the speed of light). Thus we can interpret these properties as properties of an electromagnetic theory, but without committing to identifying `purely electric' or `purely magnetic' properties, which could only be defined through a choice of complex structure---and we do not need to make such a choice in the common core theory. A change of complex structure changes the putative `purely electric' or `purely magnetic' properties.

Thus the specific structure that a model adds to the common core theory is a choice of complex structure on the plane. This choice determines what properties count, in a given model, as `purely electric' or `purely magnetic'.

The common core theory itself only knows about electromagnetic properties and excitations, i.e.~excitations of a complex vector field that are transverse, propagate at the speed of light, and satisfy a dynamical equation that (for any choice of complex structure) is equivalent to the Maxwell equations.

This clarifies the sense in which a common core ontology `says less' than each of the models. Nevertheless, we see that the specific structure that is added to the common core theory to get a specific model or, alternatively, that is ``erased'' from a set of models to get the common core theory, is not a matter of `deleting degrees of freedom' or `describing the same physics with fewer variables'. Rather, what the specific structure comes down to is the specification of some choice out of a number of {\it possibilities allowed by the common core theory}: this choice then bears on how the theory is interpreted. 

That we here speak of `electromagnetic properties' without a specification, in the common core theory, of `purely electric' or `purely magnetic' properties, may sound familiar. Indeed, regardless of dualities, the Lorentz transformations map electric and magnetic fields onto each other under a change of coordinates (and this symmetry of electric and magnetic properties was of course the starting point for Einstein's formulation of special relativity).

As \citeauthor{read2016interpretation} (2016:~p.~224) has noted, this is analogous to sophisticated substantivalism in the discussion of the hole argument. Thus considered, sophisticated substantivalism is a balancing act that (i) keeps the points of a manifold in the ontology of a spacetime theory, i.e.~spacetime points are genuine entities, while (ii) it denies in some sense their individuality, by denying that spacetimes that are related by a diffeomorphism can differ solely by the properties that are assigned to the same spacetime points.

By analogy, an internal interpretation of the common core of electric-magnetic duals is a balancing act that (i) keeps electric and magnetic properties of fields in the ontology, i.e.~electric and magnetic properties are genuine properties, while (ii) it denies in some sense their individuality, by denying that electromagnetic fields that are related by a duality transformation can differ solely by the properties that are called electric and the ones that are called magnetic. 

This broadly aligns with \cite{dewar2019sophistication}'s account of sophistication. In order to interpret putatively isomorphic models as physically equivalent, instead of moving to a reduced formalism, we formulate a new theory (here, the common core theory) where the models are distinct but are rendered isomorphic. Then we interpret that theory anti-quidditistically, i.e.~we deny that two different possibilities can differ just by a permutation of fundamental (here, electric and magnetic) properties.\footnote{For the doctrine of (anti-)quidditism, see \cite{black2000against,lewis2001}. Note that the technical detail of sophistication as discussed here does not precisely match onto either of  \cite[p.~502]{dewar2019sophistication}'s two forms of sophistication,
i.e.~external and internal (for an explication of these two forms and a critique of external sophistication, see \cite[pp.~324, 332]{martens2021sophistry}). We leave a detailed comparison with external and internal sophistication for the future.}
(For more on the above discussion, see the FAQs in Section \ref{sec: faq}.)

\section{Dualities in Quantum Field Theory and Gravity}\label{DQFTG}

The previous Chapter gave an overview of major examples of dualities that can be formulated using elementary quantum mechanics, statistical mechanics, and classical field theory. 

This Chapter will focus on more advanced examples, where `advanced' here means that the examples rely on more technically demanding theories such as quantum field theories, string theory, and gauge-gravity duality. Section \ref{sec: EMQFT} discusses \textbf{electric-magnetic} duality in quantum field theory. This first topic naturally builds on electric-magnetic duality in classical field theory, discussed at the end of previous Chapter. Also, this example allows us to begin introducing some important and interesting features of dualities in these more sophisticated examples.

Section \ref{sec: stringduality} then goes on to discuss dualities in string theory, with a particular emphasis on \textbf{T-duality}, i.e.~the duality between compact dimensions of small and large radius. We focus on string theory both because of the rich conceptual tapestry that string dualities reveal, and also because string dualities were a major starting point for the recent discussion of dualities in the philosophical literature. There are two reasons for this: First, dualities are prevalent in string theory. Second, dualities are equivalences between rather different models. These two facts together lead to such an unexpected web of dualities that the need for conceptual work became inevitable. 

Finally, in Section \ref{sec: bulkads}, we discuss gauge-gravity duality, where we focus on its most developed example: namely, the holographic \textbf{AdS-CFT} duality. This duality is crucial both because it lies at the heart of much recent work on the non-perturbative formulation of quantum gravity, and because it enables us to discuss issues related to spacetime emergence and its status with respect to duality relations, bringing together two of the most important debates in current philosophy of physics and philosophy of quantum gravity. 

\subsection{Electric-Magnetic Duality in Quantum Field Theory}\label{sec: EMQFT}

This Section discusses electric-magnetic duality in classical and quantum field theory: and we use it to illustrate the themes of {\it hard-easy} and {\it elementary-composite}, as well as the {\it predictive and heuristic power} of dualities (see Section \ref{2.2}). Indeed, one key reason why electric-magnetic duality is important in a quantum field theory is because it may enable the exact prediction of the lowest-energy states of that theory. Namely, if a theory is self-dual under electric-magnetic duality, the number of electrically charged states must be the same as the number of magnetically charged states with the same mass and spin. Thus, given the knowledge of the electric states, one can use the duality to find the magnetic states. This is significant, because the magnetic states are often {\it solitonic states} that correspond to non-linear solutions of the equations at strong values of the coupling. These states cannot be found by using perturbation theory around the vacuum at weak coupling (see the discussion of Eq.~\eq{eq:pert} in Section \ref{2.2}), and so the presence of a duality is a great help towards determining them. Thus the hard-easy theme is heuristically powerful, because it offers a window into non-perturbative physics. 

\subsubsection{Particle-soliton duality in three dimensions}\label{Psd3D}

Before we discuss electric-magnetic duality in quantum field theories, it will be useful to discuss a simpler case: namely, the Maxwell theory in three dimensions (two dimensions of space and one of time), coupled to a complex scalar. This theory is often called the three-dimensional {\bf Abelian Higgs model}, where the complex scalar field plays the role of the Higgs field in an Abelian theory, i.e.~the gauge group is U(1). This example is of great importance for condensed matter systems, and its behaviour gives a blueprint for the more advanced dualities in quantum field theory (e.g.~quark confinement) and string theory. In particular, it illustrates the theme, from Section \ref{2.1}, of {\it elementary-composite.}

The leading idea is going to be that, in models where the gauge field is coupled to a scalar, there are {\it soliton solutions that are magnetically charged} and that are related by duality to non-solitonic, i.e.~particle-like, solutions that are electrically charged. (In line with our theme, we will interpret the solitonic solutions as composite, and the particle-like solutions as elementary.)

A {\bf soliton} is, in short, a solution of the (non-linear) equations of motion that has finite energy and is spatially extended, i.e.~it is localized within a finite region, and it is topologically stable. For example, solitary water waves of exceptional stability, such as waves of translation, are solitons. 

Our leading idea thus combines electric-magnetic duality with the interchange of particles and solitons. These two exchanges are required because, while electric-magnetic duality does {\it not} obtain for models with scalars, when combined with the exchange of a set of solutions (the particle solutions) with another set of solutions (the soliton solutions), one can get an electric-magnetic duality or a quasi-duality that exchanges particles and solitons.

We will first discuss the idea of particle-vortex duality and the role of vortices in generating a phase transition. Then we will discuss the duality between the topological and Noether currents, associated with solitons. (In this Section, we discuss quasi-dualities that are valid only by approximation. There will be no confusion in our calling them `dualities'.)\\
\\
{\it Vortex solutions.} In three dimensions (i.e.~two dimensions of space and one of time), the solitons we are interested in are {\bf vortices}, which are topological defects on the plane, typically defined by a singularity of the complex scalar field at the vortex's centre. The word `vortex' suggests that there is a rotational symmetry around this centre: which, for a single vortex, is indeed present. The topological character of the vortex is seen in the phase of the scalar field as we go around the vortex: it jumps by an integer $n$. This integer gives the magnetic flux of the vortex in terms of the inverse electric charge, and it satisfies the Dirac quantisation condition, Eq.~\eq{Diracq}.\footnote{For more details on vortex solutions, see \citeauthor{weinberg2012classical} (2012:~pp.~40-41, 45-47) and \citeauthor{manton2004topological} (2004:~Chapter 7). For more on particle-vortex duality, see \citeauthor{zee2010quantum} (2003:~Chapter VI.3).} \\
\\
{\it The Berezinskii–Kosterlitz–Thouless transition.} The statistical mechanics of systems of many vortices involves a phase transition. To understand why, note that, at high temperatures, the vortices are dissociated due to the thermal fluctuations. Thus the system has no particular order: it is in general disordered. However, at low temperatures, vortex-anti-vortex pairs can form and remain stable due to their binding energy (where the anti-vortices have negative magnetic flux). This formation is called the {\it condensation} of the vortices in pairs. Thus we can get a topologically ordered phase, where all the vortices are condensated in magnetically neutral pairs. The transition from one phase to the other is the Berezinskii–Kosterlitz–Thouless transition.\footnote{See \citet{kosterlitz1972long,kosterlitz1973ordering} and \citet{berezinskii1971destruction,berezinskii1972destruction}.} 

As we will discuss below, this idea of the condensation of vortices has been proposed in quantum field theory as a mechanism for confinement, through the condensation of monopoles in the vacuum.\\
\\
{\it Noether and topological currents.} We have discussed that vortices and other solitons usually have topological fluxes and charges. These are like ``winding numbers'', i.e.~they are integers that characterise topologically non-trivial properties of the magnetic field around a circle or closed surface. They can usually be obtained by integrating a {\it topological current} over a circle or closed surface surrounding the soliton.

This topological current has two important properties: (i) it is {\it not} a Noether current for some continous symmetry; (ii) it is magnetic, because it does not couple to the gauge field (as an electric, Noether, current would do, through a coupling $AJ_{\tn{Noether}}$), but to the Hodge dual field. (Recall that the electric flux is given by the electric field across a surface, which is contained in the Faraday tensor: while the magnetic flux is given by the magnetic field across a surface, which is contained in the Hodge dual of the Faraday tensor.) This explains why these kinds of dualities usually require the exchange of both elementary and solitonic solutions, {\it and} of electric and magnetic charges.

Electric-magnetic duality can be used to simplify the description of these magnetically charged solitons, thus aligning the {\it elementary-composite} theme with the {\it easy-hard} theme: the topological current associated with them can be mapped to a Noether current that couples electrically to a (dual) gauge field. The Noether current is the current corresponding to the U(1) symmetry of this dual gauge field.

For example, three-dimensional vortices have the following current associated with them:
\bea\label{vortex}
J^\l_{\tn{vortex}}=\frac{1}{2\pi}\,\e^{\m\n\l}\pa_\m\pa_\n\chi\,,
\eea
where $\chi$ is the phase of the complex scalar field around the vortex.\footnote{Despite the fact that an antisymmetric tensor is being contracted with a symmetric tensor, the vortex current is non-zero, because the phase is not single-valued.}

The above current is related by a Hodge-type (quasi-) duality to the Noether current of the scalar field:
\bea\label{Noetherc}
J_\m^{\tn{Noether}}=e\,\mbox{Im}\,(\f^*\partial_\m\f)\,,
\eea
which couples in the usual way to the gauge field, while the vortex current couples to the Hodge dual of the field. 

This ability to dualize soliton currents, and the gauge fields they couple to, into dual Noether currents coupled to dual gauge fields, opens the possibility of learning about the solitonic phase of a model by studying its dual, non-solitonic, phase. This is reminiscent of the Ising model, where Kramers-Wannier duality allowed the study of the high-temperature phase by dualizing it into the low-temperature phase.

\subsubsection{Monopoles and confinement of colour charge}\label{Mccc}

In this Section, we briefly illustrate the heuristic power of dualities, harnessed in 1975 by 't Hooft and Mandelstam to propose a mechanism for quark confinement. The mechanism involves the condensation of monopoles, and is suggested by considering the electric-magnetic quasi-dual of the confinement of the magnetic field in a superconductor to the interior of vortices. Just as, in a superconductor, the magnetic field outside the vortices is screened by the Meissner effect, due to the condensation electrons into Cooper pairs: so they proposed that, in a non-abelian theory, the electric colour charge is screened by the condensation of magnetic monopoles. Furthermore, the discussion of the properties of magnetic monopoles as soliton solutions of classical field theories will pave the way to discussing, in Section \ref{emdsym}, electric-magnetic duality in quantum field theory.

Thus this Section takes three steps: it first discusses the confinement of magnetic charge in a superconductor, then it discusses the confinement of electric (colour) charge in a dual superconductor, and then it discusses magnetic monopoles.\\
\\
{\bf Confinement of magnetic charge in a superconductor.} Recall the Meissner effect in a superconducting material, where the magnetic field is expelled from the interior of the superconductor. This is due to surface currents that shield the magnetic field in the interior, and are produced by the condensate of Cooper pairs in the superconductor. 

However, the magnetic field can be non-zero inside a {\it vortex tube} in the interior of a superconductor: so that it is, in effect, {\it confined} to the interior of magnetic flux tubes between pairs of magnetic charges with opposite charge. \citeauthor{nielsen1973vortex} (1973:~p.~45) described this phenomenon using solutions of the four-dimensional Abelian Higgs model (three dimensions of space and one of time; recall that Section \ref{Psd3D} discussed the Abelian Higgs model in two dimensions of space and one of time). They outlined a mechanism that keeps magnetic charge in a superconducting material confined in the interior of elongated vortex tubes with opposite magnetic charges at their ends, thereby securing magnetic charge conservation. This confinement of magnetic charge in the interior of vortex tubes leads in to the idea of confinement of electrically charged quarks, in a model of ``dual superconductivity'':\\
\\
{\bf A dual superconductor: confinement of colour charge from condensation of monopoles.} Subsequently, \citet{hooft1975gauge} and \citet{mandelstam} proposed that the mechanism of quark confinement could be understood as the {\it electric dual} of the confinement of the magnetic field in the interior of magnetic flux tubes in a superconductor. The flux tube is electric, rather than magnetic, so that the electric colour charges, i.e.~the quarks, are attached to the ends of the string. The analogue of the superconducting background that confines the magnetic charge to the interior of the vortices, i.e.~the condensate of electrons in Cooper pairs, is a {\it condensate of magnetic monopoles} in the vacuum, that confines the electric colour charge, in a kind of electric {\it dual Meissner effect}.

There are of course important disanalogies between the confinement of magnetic charge in a superconductor and the confinement of colour charge in QCD. Importantly, the superconductor model is abelian, while colour charge is non-abelian. Furthermore, the 't Hooft-Mandelstam mechanism itself requires dualizing the Higgs field so as to get to a disordered phase with a dual disorder parameter.

Explaining colour charge confinement remains a major open problem in the Standard Model, and it is possible that its explanation is not given by a single mechanism like the 't Hooft-Mandelstam mechanism. In fact, it is very likely that an adequate explanation will also involve other non-perturbative effects.

Nevertheless, there is evidence, from lattice models, that the 't Hooft-Mandelstam mechanism does play a role in confinement. Furthermore, it is realized in certain supersymmetric quantum field theories, where it can be studied using electric-magnetic duality. Since quark confinement is such a major open problem in physics, the possibility of explaining it using duality is sufficient motivation to study electric-magnetic duality in non-abelian gauge theories.\\
\\
{\bf Magnetic monopoles.} The previous study of electric-magnetic duality requires the discussion of magnetic monopole solutions, which are the carriers of {\it magnetic charge}. Monopoles in non-abelian gauge theories share key formal properties with the vortices in the previous subsection. Like vortices, they have an associated {\it topological current} (see Eq.~\eq{vortex}), whose integral is an integer that is topologically invariant (namely, the degree of a certain map between spheres), and their charge is quantised according to Dirac's quantisation condition.

Although, from a distance, monopoles in non-abelian gauge theories look like Dirac monopoles, they are in fact very different. For, unlike Dirac monopoles, they have a smooth core. Also, unlike Dirac monopoles, the mass of a monopole in a non-abelian gauge theory is predicted by the theory, and satisfies the Bogomol'nyi bound that relates the mass, $M$, and its magnetic charge, $g$, as follows:
\bea
M\geq|vg|\,,
\eea
where $v$ is the vaccum expectation value of the Higgs field. Thus magnetic monopoles are very natural solitonic solutions of non-abelian quantum field theories.

Note that, in general, there is no duality between the electrically charged particle excitations, or states, of a quantum field theory and its magnetic monopole states. Thus a general quantum field theory does not satisfy electric-magnetic duality.

However, in the next Section we will discuss important examples of quantum field theories, especially those involving supersymmetry, that do enjoy a version of electric-magnetic duality. In such cases, duality is a powerful tool that allows us to study the theory's spectrum. 

\subsubsection{Electric-magnetic duality in supersymmetric Yang-Mills theory}\label{emdsym}

In this Section, we will discuss two examples of electric-magnetic duality for quantum field theories. Both concern a supersymmetric version of Yang-Mills theory with gauge group SU(2), although we will not require details about supersymmetry. The first example is a theory that is mapped onto itself under electric-magnetic duality, i.e.~it is a self-duality (see Section \ref{2.2}). The second example is a case of a (quasi-) duality that relates different (low-energy) models of a given theory.

{\it Supersymmetry} is a spacetime symmetry that extends the Poincar\'e symmetry, such that each boson has a supersymmetric fermionic partner, and vice versa. It does so by adding fermionic generators to the Poincar\'e algebra, thus getting a super-Poincar\'e algebra.\footnote{A useful overview of supersymmetry is \citet{ferrarasu}. For an introduction to supersymmetric Yang-Mills theory, see \citet{dhokerphong}.} 

One first point to clarify is that, in general, supersymmetry is {\it not} itself a duality (nor is it a case of self-duality). This is because, although supersymmetry maps bosonic states and quantities into corresponding fermionic states and quantities of the same theory, it does so without preserving the structure of the state-space and the values of the quantities. Thus it is in general %an automorphism, but 
not an isomorphism of supersymmetric theories.

Because supersymmetry relates the bosonic and the fermionic states, and is encoded in the super-Poincar\'e algebra, the supersymmetry algebra contains a great deal of information about the states. In particular, it contains information about the masses, charges and spins of the theory's normal, i.e.~perturbative, particle states, which are in general electrically charged: and also about the theory's {\it solitonic states}, such as magnetically charged (monopole) states.\footnote{Recall, from Sections \ref{Psd3D} and \ref{Mccc}, that in quantum field theories, electric-magnetic duality maps the elementary electric states to solitonic magnetic states. For supersymmetric theories, this fact is also encoded in the supersymmetry algebra.}

Thus if one is looking for a theory that, as in our first example, is mapped onto itself under electric-magnetic duality, one should first check whether there are electric and magnetic states in the supersymmetry algebra that have the right quantum numbers for them to enter in an electric-magnetic duality relation. In particular, for each electric state there should be a magnetic state in the spectrum with the same mass and spin, and vice versa.

In what follows, $\mathcal{N}$ will indicate the ``amount of supersymmetry'': namely, the number of ways in which one can exchange a boson and a fermion under supersymmetry. Models with higher $\mathcal{N}$ have more supersymmetry, because there are more ways to exchange bosons and fermions. \\
\\
{\bf First example: $\mathcal{N}=4$ SU(2) supersymmetric Yang-Mills theory.} $\mathcal{N}=4$ supersymmetric Yang-Mills is the version of Yang-Mills theory with the maximum amount of supersymmetry for a theory without gravity (namely, $\mathcal{N}=4$). It has been conjectured to be {\it self-dual} under electric-magnetic duality, so that the map that maps electric states onto magnetic states {\it within the same theory} is in fact an isomorphism (and likewise for quantities), as we will next discuss.

We first give an argument to the effect that, as required by the duality, the lowest-energy states and their properties match. Then we discuss the quantities. These two steps follow the Schema's conception of duality, in Section \ref{2.1}, as a pair of isomorphisms: one for states, and one for quantities.

The lowest-energy {\it elementary particle states} of this theory are the elementary (particle) excitations of the fields. In order to count these lowest-energy states, we count the elementary excitations of the fields of the theory: (i) a massless gauge field, (ii) six scalar fields, and (iii) four spin-1/2 fields. Here, (i) and (ii) are bosonic fields, and (iii) are fermionic fields. Since the massless gauge field, i.e.~(i), has two helicity states, and each of the scalar fields, i.e.~(ii), has one state of helicity zero, there are a total of {\it eight bosonic} elementary states, obtained by perturbative quantization.\footnote{We here follow the terminology used in physics, where, in the context of counting elementary one-particle states, one only counts the independent states which can be used as a basis for the relevant Hilbert (sub)space. Thus one is counting the quantum numbers that label states in irreducible representations of the symmetry groups of the model. After this counting, one has to verify that the bijection is in fact an isomorphism: and we do this in the following step.} 
This amount, i.e.~eight bosonic states, equals the number of fermionic states, i.e.~(iii), because each of the four fermions has two helicity states: and so, the total number of fermionic states is also eight (and this is itself an expression of the supersymmetry of the theory). Thus there are a total of {\it sixteen lowest-energy elementary particle states}, i.e.~states that correspond to elementary excitations of the bosonic and fermionic fields. 

To have electric-magnetic duality, these elementary particle states must match the lowest-energy {\it solitonic states}: which, as we discussed above, can be found from the supersymmetry algebra. These states are similar to the monopoles we discussed in Section \ref{Mccc}: they are not elementary excitations of the fields, but rather excitations of the fields around (usually) non-linear solutions of the equations of motion. As it turns out, for each particle state there is a solitonic state with the same mass and spin, but with magnetic, rather than electric, charge. 

Thus it follows from the supersymmetry algebra that this theory indeed has a chance of mapping onto itself under electric-magnetic duality, which in this context is called {\bf Montonen-Olive duality} (see \citet{montonen1977magnetic}). Namely, the duality relates the electric and magnetic states of lowest energy in the correct way that we have just discussed. We say `has a chance', because a duality must also preserve all of the theory's quantities (see Section \ref{2.1}, especially Figure \ref{obv2}).

In fact, the partition function and the free energy also transform in the correct way: which suggests that, at least for the lowest-energy states, the electric-magnetic duality of the $\mathcal{N}=4$ Yang-Mills theory is true.\footnote{These lowest-energy states are called BPS states, and for these states the evidence points to the existence of a duality. However, the duality involves constructing non-perturbative magnetic states, and there is no rigorous mathematical proof that it can be done. Also, the duality needs to be demonstrated for all other quantities.}

However, much more needs to be done for a full demonstration of electric-magnetic duality, even on the physics level of rigour. For the higher-energy states need to respect the duality as well.\footnote{There are several important subtleties, such as for example the role of the gauge group SU(2) under the duality. For a discussion, see \citeauthor{deharobutterfieldOUP} (2024:~Chapter 7).}

The electric-magnetic duality of $\mathcal{N}=4$ Yang-Mills theory illustrates our general themes of {\it hard-easy} and {\it elementary-composite} as follows:

About {\it hard-easy}: as we have mentioned, elementary electric one-particle states are related, by electric-magnetic duality, to magnetic solitonic states. (And this is like in previous examples, in Section \ref{sec: EMQFT}.) Here, the magnetic states are non-perturbative, because the weak-coupling regime of the electric states is mapped, under the duality, to the strong-coupling regime of the magnetic monopole states. This is because magnetic monopoles couple to the Lagrangian with their magnetic charge $g$, rather than with the electric charge $e$. And since these charges satisfy the Dirac quantisation condition, i.e.~Eq.~\eq{Diracq}, the duality maps the perturbative regime of small $e$ to the non-perturbative regime of large $g$, and vice versa. Thus a problem that is easy to solve (in perturbation theory) for the electric states and quantities, is hard to solve for the dual, and vice versa.

About {\it elementary-composite}: as we discussed above, magnetic monopoles in $\mathcal{N}=4$ Yang-Mills theory are not elementary one-particle excitations of the fields, but are rather solitonic excitations, which are associated with topological configurations of gauge fields. They are topologically non-trivial in the sense that they cannot be continuously deformed to the trivial vacuum solution, which is the solution about which one expands in perturbation theory. This topological nature means that the solutions are extended in space, and are not localised. Their stability is due to topological reasons, rather than to local currents and conservation laws. Thus, just like the vortices of the Abelian Higgs model, they are solitons with a topological, rather than a Noether, current (i.e.~analogous to Eqs.~\eq{vortex} and \eq{Noetherc}). Their topological charge is conserved and quantised, and it is not associated with a continuous symmetry and its conservation law, i.e.~it is not due to Noether’s theorem, as is the case for the symmetries of elementary states.\\
\\
{\bf Second example: $\mathcal{N}=2$ SU(2) supersymmetric Yang-Mills theory.} This theory has half the amount of supersymmetry of the $\mathcal{N}=4$ theory just discussed. The $\mathcal{N}=2$ theory does not have Montonen-Olive duality in the way outlined for the $\mathcal{N}=4$ case, because there is a {\it mismatch} between the spins of the particle and the soliton states (specifically, there is no soliton state with spin one: while, just as in the $\mathcal{N}=4$ case, there {\it is} a spin-1 particle). This means that, under a map that exchanges electric and magnetic states, the electric states are mapped to magnetic states with a different spin. Thus the electric and magnetic subspaces of the state-space are not isomorphic, nor is the state-space of the model mapped into an isomorphic state-space by a duality map.\footnote{This can also be understood in terms of the supersymmetry multiplets in which the electric and the magnetic BPS states appear. These are different multiplets: they form inequivalent representations of the super-Poincar\'e algebra, and so they are not isomorphic. For details, see \citeauthor{deharobutterfieldOUP} (2025). The original work discussed in this Section is in \citet{seiberg1994electric}. For introductions to this work, see \citet{bilal1997duality,alvarez1997introduction}.} 

Despite this mismatch between the elementary properties of the states, the values of some important quantities do approximately match. For example, the Wilsonian effective action that gives an effective description of the electric states, matches onto the Wilsonian effective action that gives an effective description of the magnetic states. Thus there is a {\it quasi-duality}, which involves a match of a subset of states and a subset of quantities that are relevant at low energies. (This match is `approximate', in the sense that it is valid at low energies, and for the range of parameters for which the Wilsonian effective action is well-defined.)

The Wilsonian effective action gives a low-energy description of the states that are relevant for given values of the parameters. One such parameter is the expectation value of the Higgs field, which plays the role of a coupling parameter, because it appears multiplying various terms in the Wilsonian effective action that represent couplings between fields.\footnote{More precisely, the relevant parameter is the gauge-invariant complex number obtained by taking the expectation value of the trace of the square of the Higgs field, i.e.~$u:=\bra\Tr\f^2\ket$. However, for simplicity, we will continue to speak of `the expectation value of the Higgs field'.}\\
\\
{\it Moduli space as space of models.} In supersymmetric Yang-Mills theory, the Higgs field is a complex scalar field, and so its expectation value is a complex-valued function that depends on other parameters, such as masses, couplings, and energies of particles. Since the Higgs field is complex-valued, it spans a two-dimensional field space that is called the {\bf moduli space}, i.e.~the space of field configurations. Here, it is the space of vacuum configurations of the Higgs field, i.e.~those configurations that minimise the quantum effective potential. 

Since the Higgs field plays the role of a coupling parameter in the Wilsonian effective action, different values of the Higgs field give rise to different Wilsonian effective actions. In other words, there are regions in the moduli space around special points where the Wilsonian effective action can be written down explicitly, to good approximation within that region. The expansion that one gets in this approximation is reminiscent of a perturbative expansion in quantum field theory, which can be illustrated by Feynman diagrams, around some value of the coupling constant (usually, at zero coupling: see the discussion in Section \ref{2.2}). The difference is that here we expand the Wilsonian effective action around some expectation value of the Higgs field, and furthermore the series of terms contains an infinite number of non-perturbative (`instanton') terms that do not appear in perturbative expansions of Feynman diagrams. Nevertheless, the series is well-defined, and all the terms can in principle be calculated. This was an important, indeed epoch-making, result by \citet{seiberg1994electric}.

In each region in which we approximate the Wilsonian effective action, the action takes the form of a particular quantum field theory model. We will here say a bit more about two such low-energy models that are valid in different (partly overlapping) regions of the moduli space, namely: (1) an electric, and (2) a magnetic low-energy model.\\
\\
(1) $\,$ The {\it electric model} is the one that is valid for {\it very large} vacuum expectation values of the Higgs field. It is the supersymmetric Maxwell theory, which appears here as the massless, low-energy limit of the original SU(2) supersymmetric Yang-Mills theory after symmetry breaking by the Higgs mechanism.\footnote{For a recent discussion of symmetry breaking, see \citet{berghofer2023gauge}. Although local gauge symmetries cannot be broken, a global subgroup of the local gauge group {\it can} be broken. We thank Silvester Borsboom for discussions of this point.}\\
\\
(2) $\,$ The {\it magnetic model} is valid for vacuum expectation values of the Higgs field that are {\it near the cut-off} of the Wilsonian effective action of the electric model (and so, their values are neither very large nor very small), and it is completely different (and it is not isomorphic) to the electric model. It is supersymmetric quantum electrodynamics, i.e.~a sypersymmetric version of a spin-1/2 particle (the model also has four scalar fields, so as reach $\mathcal{N}=2$ supersymmetry). Unlike all the other supersymmetric theories we have discussed so far, this model has no vector fields. 

And yet, despite the fact that the state spaces are so different, a quasi-duality transformation maps the two models in such a way that their Wilsonian effective actions transform into each other, and their values also match in the region where the two models overlap. 

We have discussed that the moduli space of the theory is the low-energy configuration space of the Higgs field, with different regions, in each of which the Wilsonian effective action takes a different form, i.e.~with different low-energy models in each of the regions. Since in each region the system shows qualitatively different behaviour, characterised by different symmetries, different particle content and different quantities, each of which are associated with a region, it is appropriate to, by analogy with thermodynamics, think of these regions as different {\it phases} of the system.\footnote{The analogy with phase transitions in thermodynamics, such as those in the Ising model, extends further: at the critical temperature, the free energy and specific heat diverge, indicating a second-order phase transition with no latent heat. Likewise, in quantum field theories, the free energy has singularities at special critical points (or lines or regions) where the low-energy description is not valid. These singularities give the moduli space a non-trivial topology. Note that these singularities have codimension larger than zero, and so the system allows for a continuous transformation from one phase to another without encountering a phase transition. For a more precise description of the Ising model theory as a manifold, see \citeauthor{deharobutterfieldOUP} (2025).} Namely, each region has a set of well-defined quantities, some of which play the role of {\bf order parameters} for a particular phase, in that the non-zero value of that particular quantity on the states of that region indicates a characteristic property of that phase. Thus in the region where the electric model is valid, the Higgs field is a good order parameter. This region is characterised by e.g.~the breaking of gauge symmetry by the Higgs mechanism. In the magnetic region, the approximation ceases to be valid, and one needs to define a different (quasi-dual) quantity to play the role of the order parameter for that phase. Thus these other (quasi-) dual phases are characterised by other behaviours, such as quark confinement\footnote{To get quark confinement, one needs to introduce quark fields into the theory. This was done in \citeauthor{seiberg1994electric} (1994a:~pp.~41-44); see also \citeauthor{seiberg1994monopoles} (1994b:~pp.~491, 506-508).} or long-range Coulomb attraction. They go under the name of `Higgs', `confining' and `Coulomb' branches, respectively.

Besides the state of the Higgs field, the moduli space also encodes other information: the geometric quantities defined on it also encode information about the states of particles and solitons: in particular, their masses and charges. 

In our example, the moduli space is a two-dimensional space with three punctures, and it is equipped with a complex structure and a (Kahler) metric: it is a Riemann surface. The metric originates in the metric in field space with which the Wilsonian effective action is equipped. 

The quasi-duality map(s) that we discussed above are transition functions between the coordinatizations of this Riemann surface on the overlaps between open sets. What we called `regions' above thus define open sets that endow the space with a topology and, together with the coordinates, make it into a differentiable manifold. The values of the order parameters such as the Higgs field are thus coordinates on the Riemann surface, with a validity on a given region (namely, the domain of convergence of the quantities in the Wilsonian effective action, written in the coordinates in the given region).

As \citeauthor{deharobutterfieldOUP} (2025) have argued, the above view of a physical theory generalizes to many other examples in quantum field theory and string theory. They have argued that this gives, more generally, a view of physical theories that they have dubbed the {\it geometric view of theories}, and which generalizes the semantic conception. We will return to this view of physical theories in Chapter \ref{philiss}.\\
\\
{\bf Illustrating the themes and roles.} The Seiberg-Witten theory illustrates our themes from Section \ref{2.2} in a way that is qualitatively very similar to how our other examples of particle-soliton dualities in Section \ref{Psd3D} did.\footnote{With the appropriate differences, similar remarks apply to the $\mathcal{N}=4$ theory.}
The {\it hard-easy} theme is illustrated, because the regions where the models are valid are different and their overlaps are limited. On an overlap near a singularity where the electric model (1) becomes invalid, so that the expansion of the free energy in terms of the expectation value of the Higgs field does not converge, a quasi-duality transformation saves the day: the dual description has a convergent free energy in terms of a dual field, and so the problem becomes tractable in terms of the magnetic model (2).

The {\it elementary-composite} theme is realized in a slightly different way than in other examples we have discussed in this book. This is because we have a quasi-duality rather than a duality: namely, the models (1) and (2) are only low-energy limits of the $\mathcal{N}=2$ Yang-Mills theory, so that the latter is not a common core theory, but rather a more comprehensive theory (one could also say: a {\it successor theory} of the low-energy models: see Section \ref{2.2}). The elementary excitations of the magnetic model (2), e.g.~the magnetic monopoles, are solitons of the $\mathcal{N}=2$ theory (and they are mapped, by the quasi-duality on the overlaps, to the electric particle states of the electric model).

\subsection{T-Duality}\label{sec: stringduality}

\begin{figure}
    \centering
    \includegraphics[width=0.60\linewidth]{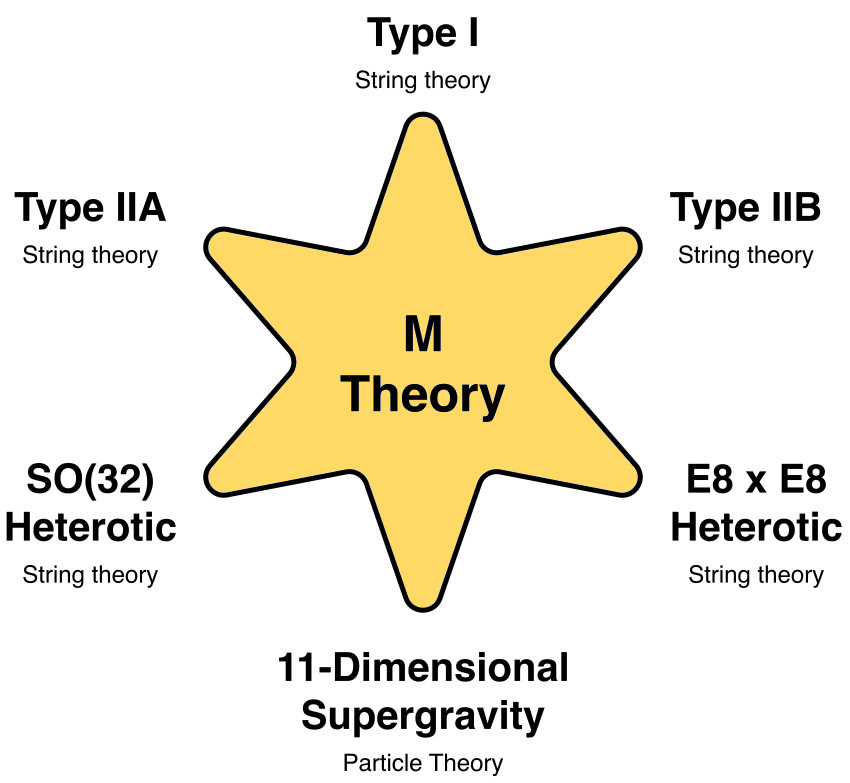}
    \caption{\small{Relations between the five superstring theories and 11-dimensional supergravity. At the centre, M-Theory is highlighted as providing a natural unification of the various string theories and with 11-dimensional supergravity. The five perturbative string theories, together with 11-dimensional supergravity, are expected to be specific limits of the more fundamental M-Theory. }}
    \label{fig:web}
\end{figure}

Having discussed an advanced example of a duality in a quantum field theory, we move on to what are arguably some of the dualities with the most far-reaching consequences: namely, string-theoretic dualities. There is a wide variety of string-theoretic examples, some aligned with field theoretic dualities, and some specific to string theory. Examples of the first kind are S-dualities, which exchange strong- and weak-coupling descriptions, thus instantiating the weak-strong theme of dualities: and gauge-gravity duality, which we will discuss in the next Section. An example of a distinctively string-theoretic duality is T-duality, which we discuss in this Section. By `distinctively string-theoretic', we mean that T-duality explicitly relies on the string-theoretic properties of the dual models. But what are string-theoretic properties? And more generally, what are the basic features of string theory? Before we dive into T-duality, we will first address this question.\\
\\
{\bf String theory and dualities.} String theory is an attempt to construct a theory of quantum gravity, i.e.~a to provide a quantum description of general relativity. Furthermore, string theory is also sometimes referred to as a theory of everything. For, in attempting to give a quantum description of gravity, it includes and unifies general relativity with the standard model of particle physics, thus describing all four fundamental forces in a single theory. The basic idea behind string theory is that the fundamental building blocks of reality are not, as traditionally thought, particles, i.e.~$0+1$-dimensional objects, but rather extended strings, i.e.~$1+1$-dimensional objects. This idea is formalized in \textit{perturbative string theory}, i.e.~the study of string theory for small values of the string coupling constant $g_s$. This theory is by now reasonably well-understood, and displays a variety of interesting dualities, of which T-duality is one. 
%Indeed, as we will see in what follows, T-duality is a paradigmatic string-theoretic duality, because the extended nature of the string is a central element of the duality. 
Finally, let us mention that, while we have so far only spoken of `string theory', there are actually \textit{five} distinct perturbative string theories. Furthermore, while these are distinct theories at the perturbative level, there is a web of dualities (see Figure \ref{fig:web}) that relates them to one another, and that is usually understood as hinting to the existence of a single non-perturbative formulation of string theory.

Nevertheless, the \textit{non-perturbative} definition of string theory is much less understood, i.e.~the formulation that is relevant to strongly-coupled gravitational phenomena like the endpoint of black hole evaporation and the recovery of the information. The theory that is supposed to encode the non-perturbative data of string theory is called \textbf{M-theory}.\footnote{It has been said that `M' here stands alternatively for `matrix', `membrane', or `mystery' (\citeauthor{witten1995string} 1995).} 
Here, the current understanding of string theory is significantly less developed: concrete formulations of (corners of) this non-perturbative M-theory are the BFSS matrix model of (\citeauthor{banks1999m} 1999) and the AdS-CFT correspondence (\citeauthor{maldacena1999large} 1999;  \citeauthor{witten1998ads} 1998), which will be discussed in the next Section. However, the main hint towards the existence and nature of M-theory is given by the web of dualities that are believed to obtain between the five perturbative superstring theories (more precisely, between the five superstring theories and 11-dimensional supergravity, which would be a low-energy effective theory for M-theory), and which establishes their equivalence. This equivalence is then understood, as per the discussion in Chapter \ref{dualint}, in terms of a \textbf{successor theory}, encompassing and extending the structure of the five perturbative superstring theories.\footnote{Note that we do not think here in terms of a common core theory, since M-theory is supposed to be more general than perturbative string theory, rather than simply encoding its duality invariant content.} Insofar as this theory exists, it is natural to think of it as the sought-after M-theory. Interesting as these questions are, both for the physics and the philosophy of dualities in string theory, in this Chapter we will confine ourselves to perturbative string theory, and explore the basics of T-duality there.\\
\\ 
{\bf Perturbative string theory.} We first briefly introduce the basic features of perturbative string theory.\footnote{For a detailed introduction to string theory, see, e.g.~\citeauthor{becker2007string} (2007). An introduction and discussion of various philosophical issues is given in \cite{wuthrich2025out}.} 
At the perturbative level, string theory is naturally understood in terms of so-called \textit{$\sigma$-models}. A $\sigma$-model, in the present context, is a two-dimensional field theory, whose fields are maps from the two-dimensional manifold that is interpreted as the \textit{worldsheet} swept out by the string as it propagates, to a higher dimensional manifold, called \textit{target} manifold or spacetime, which is the ambient space where the string moves. In string theory, the target manifold is a ten-dimensional spacetime that is a solution of supergravity, i.e.~the supersymmetric extension of Einstein’s general relativity. This spacetime gives the background against which one sets up the perturbative expansion constituting perturbative string theory, while the fields of the $\sigma$-model describe how the two-dimensional worldsheet is embedded into the ten-dimensional target space. Hence, string theory describes, via a $\sigma$-model, the dynamics of strings propagating in a ten-dimensional spacetime. In particular, the action of the two-dimensional field theory defining the $\sigma$-model is the \textit{Polyakov} action, which reads as follows:
\bea\label{eq:poly}
\mathcal S_{\tn{Poly}} \left[h,X\right]= \frac{1}{\ell_s^2} \int \dd^2 \sigma\, \sqrt{-h}~h^{\alpha\beta} \partial_\alpha X^\mu \partial_\beta X^\nu g_{\mu\nu} (X^\rho)\,,
\eea
where $\ell_s$ is the string length, i.e.~the length scale for the extension of the string, $\sigma^0 = \tau$ and $\sigma^1= \sigma$ are the world-sheet coordinates, $h^{ \alpha \beta}$ and $h$ are respectively the inverse metric and the determinant of the world-sheet metric $h_{\alpha\beta}$, which describes the geometry of the world-sheet. $X^ \mu \left( \sigma \right)$ is a map between the string world-sheet and the target space, i.e.\ the spacetime in which the string propagates, while $g_{ \mu \nu}(X)$ is both the coupling function of the string interactions and the metric on the target space.

Here, the embedding of the worldsheet into the target spacetime is given by the $X^ \mu$ fields, each of which gives the position of the string in one of the dimensions of the target space. Hence, we have ten $X^ \mu$ fields in string theory, one for each dimension of the spacetime where the string propagates. This action has conformal symmetry under conformal transformations of the metric $h_{\alpha\beta}$, and it describes a two-dimensional \textit{conformal} field theory. This observation is important, since the conformal symmetry of the Polyakov action is not preserved upon quantization, giving rise to what in quantum field theory is called an \textit{anomaly}, i.e.~a classical symmetry broken by quantization. It is required that this type of anomaly cancels, i.e.~one adds terms to the action such that they cancel the contribution from the term creating the anomaly. In string theory, cancelling the conformal anomaly is equivalent to requiring that the target space is governed by the Einstein equations (plus quantum corrections) in their appropriate extension to cover ten-dimensional supergravity, and hence it is crucial to string theory’s claim to be a theory of gravity, since it implies that it actually describes gravity in the classical limit. However, cancelling the conformal anomaly requires (once supersymmetry is included) ten $X^ \mu$ fields, i.e.~$\m=0,\ldots,9$, so that the target spacetime is ten-dimensional.\footnote{Absent supersymmetry, one requires 26  $X^ \mu$ fields, i.e.~$\m=0,\ldots,25$, hence a 26-dimensional target space.} 
This is the origin of the requirement, in string theory, that there are extra dimensions beyond the four dimensions of relativistic spacetime. \\
\\
{\bf Compactification.} At first sight, the requirement that there are extra dimensions poses a phenomenological problem for string theory, for we seem to be living in a four-dimensional, not a ten-dimensional, spacetime. One way out of this predicament is to postulate that these extra dimensions are \textbf{compact} (in the physicists', not the mathematicians', sense). To see what physicists mean by `compactness', think of a five-dimensional universe in which one of the dimensions has the topology of a circle $S^1$. We can describe phenomena whose characteristic length scale is much larger than the circle’s radius by using an effective field theory in a four-dimensional spacetime with some additional degrees of freedom (encoding the residual physics coming from the fifth dimension), thus in effect removing the fifth dimension or, at least, making it invisible at large distance scales. We say that the fifth dimension has been {\it compactified}, and the mechanism for this compactification is the Kaluza-Klein mechanism, after the two physicists who initially introduced it (\citeauthor{theodor1921unitatsproblem} 1921; \citeauthor{klein1926quantentheorie} 1926). In string theory, the compactified dimensions are six rather than one, and their topology and geometry are much more complicated than a circle: a well-studied example is that of compactifications on six-dimensional Calabi-Yau manifolds, which give rise to supersymmetric extensions of the standard model of particle physics. Nonetheless, the basic idea is the same as the five-dimensional example we have discussed.

In the context of T-duality, it will be enough for us to think of a ten-dimensional spacetime that is a solution of perturbative string theory, with a single dimension compactified on a circle $S^1$, without having to look at the complexities of Calabi-Yau compactifications. We highlight two properties that are especially relevant for the description of a string propagating in the compact dimension: the momentum $P$ of the string and its winding number $n$. The momentum is the usual quantum mechanical observable that we encountered earlier in this book, and given that we are studying propagation along a compact dimension, upon quantization it will have a discrete spectrum. By contrast, the winding number is a purely string-theoretic property, since it counts the number of times that the string winds around the compact dimension: a property that no particle could have, since it lacks extension in space and hence it cannot be wound around anything. Also the winding number of the string has, upon quantization, a discrete spectrum. Finally, we also need to keep track of the radius of the circle, i.e.~$R$.\\
\\
{\bf Invariance of the spectrum under T-duality.} Recall, from Section \ref{2.1}, that a duality maps states between models, such that the values of the quantities match. Here, we will show that under the exchange of quantum numbers $n$ and $m$, the masses of the states indeed remain invariant. Thus consider the mass formula for a string propagating in the compact dimension. It takes the following form:
\bea\label{eq:mass}
M^2 = \frac{n^2}{R^2} + m^2R^2 + \mbox{oscillations}\,,
\eea
where $n$ is the quantum number characterizing momentum modes, $m$ is the quantum number characterizing winding modes, $R$ is the radius of the compact dimension i.e.~the circle, and `oscillations' refers to the oscillation modes of the string, which we suppress because they are not relevant for our argument. We also follow the widespread convention of setting the string length equal to one, i.e.~$l_{\sm s}=1$.

Notice in particular the first two terms, given by the momentum and winding numbers. We can see that we can get a mass equivalent to that in Eq.~\eqref{eq:mass} if we carry out the following change of variables:
\bea\label{eq:tduality}
m \leftrightarrow n, \  R \leftrightarrow \frac{1}{R}\,.
\eea
In other words, the mass formula for a string propagating along a compact circular dimension is invariant under a transformation that exchanges the  quantum numbers of winding $m$ and momentum $n$, and changes the radius of the compact dimension from $R$ to $\frac{1}{R}$. This equivalence illustrates a more general fact in string theory: namely, pairs of solutions with compact dimensions, whose radii have reciprocal values, are equivalent. This equivalence extends to all other quantities, and it is what we call T-duality.

Thus we have an equivalence between a string theory defined in a spacetime with a compact dimension with radius $R$ and one defined in a spacetime with a compact dimension with radius $\frac{1}{R}$. This implies that, when we have a model in a spacetime that has a small radius and hence a singular compact dimension, we can, by an appropriate change of variables, i.e.~by exchanging winding and momentum quantum numbers, get a model where the compact dimension has a large radius and is non-singular. Note that this equivalence is only the simplest case of T-duality: in general, a spacetime can have any number of compact dimensions, on each of which we can apply a T-duality map. Furthermore,  related to T-duality is a more general duality called \textit{mirror symmetry} (\citeauthor{hori2000mirror} 2000), which not only concerns the size of the compact dimensions, but also their topology, and which has proven fruitful, beyond its string-theoretic applications, in pure mathematics (\citeauthor{kontsevich1995homological} 1995; \citeauthor{strominger1996mirror} 1996).\footnote{Indeed, \cite{strominger1996mirror} arguably provides the clearest connection between mirror symmetry and T-duality: there, mirror symmetry is conjectured to be realized by a T-duality map on certain fibrations of the Calabi-Yau manifolds modelling the compact dimensions of two string theories.}\\
\\
{\bf Roles of T-duality.} An interesting feature of T-duality is its role within the web of string dualities that are taken to point to the existence of M-theory, which we discussed at the beginning of this Section. Going back to Figure \ref{fig:web}, we see that T-duality relates Type IIA and Type IIB string theories. This equivalence is an instance of what we described in Chapter \ref{dualint} as the \textbf{heuristic role} of dualities, i.e.~dualities' role as clues towards a more fundamental and encompassing theory which is expected to supersede the two dual models, as their \textbf{successor theory}. T-duality is in fact part of the motivation for developing M-theory, which is a fundamental, non-perturbative formulation of string theory: and note that, as one expects from a successor theory, M-theory indeed goes beyond the common core of T-duality, since it encompasses all five superstring theories and $11$-dimensional supergravity (thus it goes well beyond the T-dual models). Furthermore, insofar as M-theory is well-understood, its basic structure is expected to go beyond that of superstring theory, which might only be recovered in a specific limit of M-theory. Thus T-duality is used heuristically to find clues about M-theory. For example, by studying how T-duality maps the Type IIA model as its coupling increases, one hopes to get clues about some of the corners of M-theory. By allowing the coupling of Type IIA to be large, we get the corner of M-theory describing the strong coupling behaviour of Type IIA. If we compactify this corner of M-theory on a torus, and apply T-duality to one of the two circles which make up the torus and then decompactify the compact dimensions, we get the corner of M-theory describing the strong-coupling behaviour of Type IIB string theory, i.e.~F-theory (\citeauthor{vafa1996evidence} 1996).\footnote{For details about the relation between M-theory and F-theory sketched above, see \citeauthor{Witten_1996} (1996). For a review of F-theory see \citeauthor{ceccotti} (2010); \citeauthor{weigand2018f} (2018). For a philosophical discussion of F-theory, see \cite{cinti2024time}.} 
While a detailed introduction to F-theory goes beyond the scope of this book, its discovery and relation to M-theory through T-duality is a powerful reminder of the heuristic power of dualities. \\
\\
{\bf The ontology of T-duals.} Before moving on, let us comment on an influential argument about the ontology of T-dual models due to \citeauthor{huggett2017target} (2017). \citeauthor{huggett2017target} (2017) argues that T-duality gives rise to a distinction between the physical space of string theory and its target space. Roughly, the idea is that, since the size of the compact dimensions is not invariant under the T-duality transformation, the extra compact dimension should not count as part of the physical space where a string propagates. Support for this idea is drawn from a thought experiment by \citeauthor{brandenberger1989superstrings} (1989) that envisages a procedure for determining the size of the extra dimension. This thought experiment suggests that any measurement  of the extra dimension that we could make will return the result that it is large, regardless of whether the extra dimension of the target space is large or small (with a correspondingly inverse size for the space of winding modes of the string). In the language of the Schema, the argument says that the compact extra dimension is not part of the common core of T-duality: more precisely, the information about the size of the extra dimension is not in the common core, but the fact that an extra dimension exists is true in both models and is part of the common core, possibly together with e.g.~the topology of the extra dimension.

T-duality is an interesting case of a duality transformation in perturbative string theory. However, it does not have any explicit information about the non-perturbative regime: at best, we might want to {\it extrapolate} T-duality to that regime, but it does not describe it directly. In the following Section, we will look at an attempt to define the non-perturbative behaviour of quantum gravity, and string theory in particular, in terms of a duality relation: the AdS-CFT correspondence.

\subsection{AdS-CFT and Bulk Reconstruction}\label{sec: bulkads}

In this final Section, we will discuss what is arguably the most sophisticated example of a duality that we will encounter in this book: AdS-CFT (\citeauthor{maldacena1999large} 1999; \citeauthor{witten1998ads} 1998), and in particular bulk reconstruction. Given the complexity of this topic, our treatment will be schematic. Nonetheless, it should be enough for the reader to gain a basic understanding of the issues involved, sufficient to both understand the ensuing philosophical discussion, and to approach the physics and philosophy literature on bulk reconstruction in AdS-CFT.\footnote{For detailed introductions to AdS-CFT, see \citeauthor{aharony2000large} (2000); \citeauthor{ammon2015gauge} (2015). A philosophical introduction is in \cite{de2016conceptual}. For bulk reconstruction, see \citeauthor{harlow2018tasi} (2018). Philosophical issues related to bulk reconstruction are discussed, e.g.~ in \citeauthor{bain2020spacetime} (2020); \citeauthor{bain2021rt} (2021); \citeauthor{cinti2024beyond} (2024). Metaphysical issues related to these constructions are discussed, e.g.~ in \citeauthor{jaksland2021entanglement} (2021); \citeauthor{cinti2021humeanism} (2021).}

We begin by clarifying some of the terminology and basic notions that we will use throughout this Chapter. First of all, we introduce AdS-CFT: 
\begin{itemize}
    \item[] {\bf AdS-CFT}: the (conjectured) duality between quantum gravity in (asymptotically, locally) Anti de Sitter (AdS) spacetime in $n+1$ dimensions, and a conformal quantum field theory (CFT) in $n$ dimensions.
\end{itemize}
{\it The AdS side of the duality, or `bulk'.} AdS spacetime is the maximally symmetric solution of the Einstein Field Equations with a negative cosmological constant. By `asymptotically AdS', we mean a spacetime that approaches AdS locally near asymptotic infinity, while in the interior the spacetime metric may deviate from AdS, for example by there being a black hole (for simplicity, below we will often use `AdS' for asymptotically locally AdS). 

An important feature of AdS spacetime, especially in the context of AdS-CFT, is that in AdS there is a timelike boundary, so that a beam of light travelling out to infinity along one of the spatial directions will reach a boundary; indeed, for an $n+1$-dimensional Lorentzian AdS spacetime, this boundary is an $n$-dimensional Lorentzian manifold. It is a conformally flat manifold, i.e.~it is locally Minkowski spacetime up to a conformal transformation. In other words, the AdS boundary has both spatial and temporal extension, which is the reason why we call the boundary `timelike'. We can use the boundary of AdS to gain some geometric intuition about AdS-CFT: we can think of the dual CFT as being defined on the boundary manifold of the AdS spacetime. For this reason, we will sometimes speak of a \textbf{boundary} CFT. Also, we will sometimes simply call the AdS spacetime the \textbf{bulk spacetime}, or \textbf{bulk}. These observations both explain the fact that the CFT is in one less dimension than the AdS spacetime, and also the holographic terminology for this duality: since we are mapping an $n+1$-dimensional theory into a theory defined on its $n$-dimensional boundary, much like a hologram encodes a three-dimensional theory into its two-dimensional hologram. \\
\\
{\it The CFT side of the duality.} %On the other side of the AdS-CFT-duality, we have the CFT. 
A CFT is a %special type of 
quantum field theory that is invariant under conformal transformations, i.e.~under (local) changes of scale. In other words, in a conformal field theory all the physical content of the theory is encoded in the angles between trajectories in spacetime, while lengths do not play any role, since they can always be changed under a local rescaling, and the (classical) theory is defined to be invariant under such transformations. An important point, for present purposes, is that a CFT is just a normal quantum field theory, and in particular it does not describe the gravitational force: thus it is not a theory of quantum gravity, and it is more similar to the quantum field theory of the Standard Model of particle physics (though the latter does not have conformal symmetry, while CFTs do). This explains the sense in which we can say that AdS-CFT is an instance of gauge-gravity duality: since we are mapping a theory of quantum gravity, i.e.~quantum gravity in AdS spacetime, into a non-gravitational field theory, i.e.~the boundary CFT, we are showing that gravitational and field theoretic (gauge) degrees of freedom are equivalent.\\
\\
{\bf The duality map.} We can schematically represent the duality map in AdS-CFT by the following expression:
\bea\label{eq:adscft}
Z_{\tn{CFT}} (M) = \int \mathcal{D}g \mathcal{D}\phi e^{-i\mathcal{S}[g,\phi]}\,,
\eea
which relates: (a) on the left-hand side, the partition function of the boundary CFT, defined on the boundary manifold $M$, encodes the field theory's physical data; and (b) on the right-hand side, the path integral of quantum gravity in AdS spacetime, is an informal expression for the physical content of quantum gravity in these spacetimes. Furthermore, the validity of the perturbative expansions associated with the gravity path integral, and with the CFT is controlled by the 't Hooft coupling $\lambda = g^2_{YM}N$, where $g_{YM}$ is the coupling constant of the CFT and $N$ is the rank of its gauge group, which counts its degrees of freedom.\footnote{$\lambda$ also has an interpretation in terms of the gravity theory; however, this gravitational interpretation is most naturally stated in terms of string theory: and so, for ease of exposition, we avoid it in favour of the field-theoretic one.} In particular, small $\lambda$ corresponds to a regime where the CFT perturbative expansion is reliable, while strong coupling in $\lambda$ corresponds to the regime of validity of gravitational perturbation theory. This will be relevant when we discuss how AdS-CFT illustrates the themes and types of dualities introduced in Section \ref{2.2}.

This equivalence between the CFT partition function and the gravitational path integral in AdS, if true, secures that AdS-CFT provides an isomorphism between quantities $\mathcal{Q}$ and states $\mathcal{S}$ on either side of the duality, since both notions can be defined starting from the CFT partition function/gravity path integral, Eq.~\eqref{eq:adscft}. Furthermore, since both the partition function of the CFT and the path integral of the gravity theory encode their respective dynamical information, Eq.~\eqref{eq:adscft} also secures that the duality map is compatible with the models’ dynamics, hence realizing the structure of the Schema for dualities, as introduced in Chapter \ref{dualint}.

Thus AdS-CFT has been conjectured to be a duality between quantum gravity and a conformal field theory, and in particular it shows that, at least in the context of spacetimes that are asymptotically AdS, the problem of defining a theory of quantum gravity is formally the same as the problem of identifying an appropriate CFT. Given that our current understanding of quantum field theory is much better than our understanding of quantum gravity, this duality allows us to make significant progress in our quest towards a theory of quantum gravity. In particular, AdS-CFT gives, at least in principle, a non-perturbative formulation of quantum gravity for AdS spacetimes, since the CFT is, at least in principle, non-perturbatively well-defined. We say `in principle', because the problem of giving a rigorous and general non-perturbative definition of quantum field theory is still open.\footnote{Indeed, it is part of the Clay Institute’s Millenium Problems, see \citeauthor{jaffe2006quantum} (2006) for a review.} However, we do have some control over the relevant non-perturbative physics in quantum field theory, especially in simple examples. Furthermore, there is no known reason why there should be obstructions to defining quantum field theory non-perturbatively. In other words, it seems reasonable to say that we do know the basic principles governing quantum field theory, even if we do not yet know how to formulate the theory non-perturbatively. Hence, our \textit{in principle} qualification. Note that the analogous statements for quantum gravity are indeed unknown, and in particular that in quantum gravity the limitation towards a full formulation, at present, is not mathematical like in QFT, but rather that we do not really know the basic principles governing such a theory. Hence the progress implicit in a duality like AdS-CFT.

A crucial step in identifying an appropriate CFT dual to quantum gravity in AdS is the identification of CFT quantities that are dual to local quantities in the semiclassical approximation to quantum gravity in AdS spacetime. This is necessary to ensure that the boundary CFT is actually dual to quantum gravity in AdS: without such a mapping, we would fail to have an isomorphism between the two models, in the sense of the Schema. The programme of identifying CFT expressions for semiclassical, local bulk quantities is the programme of \textbf{bulk reconstruction} (\citeauthor{harlow2018tasi} 2018). 

The crucial concept for the bulk reconstruction programme is \textit{entanglement wedge reconstruction}. Let us briefly explain what this is. In order to define the entanglement wedge of a given boundary region $R$, we first need to define the quantum extremal surface \textbf{(QES)} associated with that region. As follows:
\begin{itemize}
\item[\textbf{(QES)}] A \textbf{quantum extremal surface} $\chi$ is defined as a surface satisfying two conditions:
\begin{itemize}
\item[(i)] \textit{Homology Constraint:} given a boundary region $B$, a surface $\chi$ satisfies the homology constraint if, for $C$ a space-like hypersurface, $\chi \cup B=\partial C$, i.e.\ the union of $\chi$ with a boundary region $B$ is the boundary of some space-like region $C$. $C$ is called \textit{homology hypersurface}. 
\item[(ii)] \textit{Extremize Generalised Entropy:} the surface $\chi$ should be a surface that extremises the generalised entropy:
\bea\label{eq:sgen}
S_{\mbox{\tiny gen}} \left(\chi\right) = \mbox{ext}\left[ \frac{A(\chi)}{4G_N} + S_{\mbox{\tiny bulk}}(\chi)\right]~,
\eea
where $S_{\mbox{\tiny bulk}}(\chi)$ is the von Neumann entropy\footnote{If the total bulk state is pure, then this von Neumann entropy is the entanglement entropy between what is inside and what is outside of the entanglement wedge.} of the bulk fields contained in $\chi \cup B$ and $A(\chi)$ is the area of the hypersurface $\chi$. 
\end{itemize}
\end{itemize}

We then define the quantum Ryu-Takayanagi or HRT surface as the QES associated with the boundary region $R$, which minimises the generalised entropy. Given the notion of HRT surface, we define an Entanglement Wedge \textbf{(EW)} as follows:
\begin{itemize}
\item[\textbf{(EW)}] Let $B$ be a boundary spatial subregion of an asymptotically-AdS spacetime. The \textbf{entanglement wedge} of $B$, which we denote by $W[B]$, is the bulk domain of dependence $D[C]$\footnote{The domain of dependence $D[C]$ is the set of points with the property that any causal curve passing through one of these points must also intersect $C$.} of the homology hypersurface $C$ delimited by the HRT surface $\chi$. 
\end{itemize}

We can now introduce the crucial tool for bulk reconstruction, which is the entanglement wedge reconstruction conjecture, which says that:
\begin{itemize}
\item[\textbf{(EWR)}] \textbf{Entanglement wedge reconstruction:} all physical quantities in $W[B]$, i.e.\ the entanglement wedge of a spatial subregion $B$, are represented in the CFT by operators in $B$.
\end{itemize} 

With the notion of entanglement wedge reconstruction, we have a way to represent local, semiclassical, bulk quantities in the language of the boundary CFT. In particular, the relation between $W[R]$ and $R$ is given by a quantum error-correcting code,\footnote{See \cite{almheiri2015bulk} for the derivation of the quantum error-correcting code structure of the holographic map, due to issues having to do with locality in the bulk semiclassical approximation. See also \cite{bain2020spacetime} for philosophical discussion of these constructions.} which is required to encode the locality properties of the bulk degrees of freedom in their CFT description. While the details of this go beyond our present purposes, what is essential to the present discussion is that the Hilbert space of the semiclassical bulk gravitational theory in $W[R]$ is mapped not to the entire Hilbert space of the boundary region $R$, but only to a specific subspace, which we call the `code subspace'. This code subspace is the space of bulk semiclassical quantities and represents semiclassical gravitational degrees of freedom in the CFT. The quantum error-correcting code gives us a map between bulk semiclassical quantities and their representation in the CFT.\footnote{Strictly speaking, even this is not the case in full generality: when dealing with situations, like the black hole interior, where gravity is strongly-coupled, the error-correcting map can receive non-perturbative corrections that make the error correction approximate, and the map non-isometric, i.e.~it does not preserve inner products. In particular, in these situations we find that the fundamental quantum gravity description has fewer states than the semiclassical one, i.e.~semiclassical gravity has some redundancies that are removed in the full, exact quantum gravity theory defined via AdS-CFT; this fact is encoded in the non-isometricity of the bulk reconstruction map. See \citeauthor{akers2024black} (2024) for details.}\\
\\
{\bf Philosophical questions.} With these ideas about AdS-CFT in hand, we can look at how holography connects to more general philosophical questions. Indeed, AdS-CFT is a powerful test case for ideas about dualities. We will see its relevance for emergence in the next Chapter: here, we focus on the themes instantiated by AdS-CFT. We will focus on two aspects of AdS-CFT that are relevant to the discussion in Section \ref{2.2}: the fact that AdS-CFT is a \textit{quantum duality}, and the fact that it instantiates the \textit{hard-easy} theme of dualities. \\
\\
{\it Quantum duality.} That AdS-CFT is a quantum duality is clear from the fact that the semiclassical limits of both quantum gravity in AdS and a CFT are theories that are not equivalent in any obvious sense. In AdS-CFT, the regimes of the coupling parameter, namely the ‘t Hooft coupling $\lambda$, where semiclassical gravity in AdS and the semiclassical CFT are respectively valid, are incompatible. For semiclassical gravity in AdS is valid at {\it strong} ‘t Hooft coupling, while the semiclassical CFT is valid at {\it weak} ‘t Hooft coupling. Thus these two semi-classical limits cannot be equivalent, because each is well-defined for a different range of values of the coupling constant. This behaviour is the hallmark of a quantum duality, i.e.~a duality that is realized only at the quantum level, which displays two distinct and non-equivalent semiclassical limits.\\
\\
{\it Hard-easy.} It is clear from the previous discussion how AdS-CFT illustrates the theme {\it hard-easy}. Since the perturbative expansions of the CFT and of the gravity theory are well-defined at reciprocal values of $\lambda$, it follows that easy, perturbative calculations in the CFT correspond to difficult, non-perturbative calculations in the gravity theory, and vice versa. This property allows AdS-CFT to provide concrete insights into the non-perturbative, highly quantum dynamics of gravity, since these are mapped to perturbative calculations in the CFT, which are computationally much more tractable than their gravitational counterpart. Indeed, an important example of AdS-CFT's ability to probe non-perturbative properties of gravity is the microscopic counting of black hole entropy in string theory of Strominger and Vafa, which can be realized using holographic methods, thereby making essential use of the hard-easy properties of holography (\citeauthor{strominger1996microscopic}, 1996).\footnote{For a philosophical and historical discussion of the Strominger-Vafa microscopic calculation of the Bekenstein-Hawking entropy in string theory, see \citeauthor{de2020conceptual} (2020) and \citeauthor{van2020emergence} (2020).} Another important application of AdS-CFT to black holes, where the hard-easy theme stands out, are recent holographic calculations of the entropy curve, i.e. the Page curve, for evaporating black holes (\citeauthor{almheiri2019entropy}, 2019 and \citeauthor{penington2020entanglement}, 2020). These are crucial to the resolution of the black hole information paradox (\citeauthor{hawking1976breakdown} 1976) and the firewall paradox (\citeauthor{almheiri2013black} 2013).\footnote{See \citeauthor{harlow2016jerusalem} (2016) for a review of the physics behind these results. For a philosophical discussion of the information loss paradox, see \citet{wallace2020informationloss}. A philosophical and conceptual analysis of the firewall paradox and its resolution using AdS-CFT is in \citeauthor{cinti2021devil} (2021).} \\
\\
{\it Real-world physics?} Before concluding this Chapter, let us remark on one interesting recent development in AdS-CFT. It is often remarked that AdS-CFT is not relevant to real-world physics. The reason for this claim is that AdS spacetime has a negative cosmological constant, while the local region of the universe that we inhabit is thought to have a positive cosmological constant. Thus it is unclear how AdS-CFT could describe physics in our universe. However, despite this fact, recent results indicate the possible relevance of AdS-CFT to real-world physics, not though the gravitational physics: rather, it is the CFT side of the duality which has proven useful to real-world experiments. In more detail, we can use CFTs with holographic duals to model certain properties of strange metals, a kind of strongly-coupled condensed matter system. It has proven necessary, in order to study various properties of the strange metals, to study the system not through its CFT description, which is largely intractable, but rather using an AdS dual system, which takes the form of an electrically charged black hole.\footnote{In particular, in this context a particular form of holographic duality is used, called \textbf{semi-holography} (\citeauthor{faulkner2011semi} 2011), where only some of the field theory degrees of freedom have holographic duals, rather than all of them as in standard AdS-CFT. For a discussion of the physics and philosophy of these dualities, see \cite{cinti2025holographic}.} 
Even more interesting, various properties of strange metals described in this way have been detected, are currently in the process of being detected, in laboratory experiments. 

Does this open a window into testing AdS-CFT in the laboratory? Could these experiments on condensed matter systems tell us anything useful about quantum gravity? If so, this would be remarkable. However, more generally, it is not completely clear what the role of the gravitational dual in these experiments is: does it explain features of the strange metals? Or is it only a calculational tool? Should we be realists about it or not? Does it make sense to speak about emergence of the gravitational system from the strange metal, or vice versa, or neither? These are philosophical questions, concerned with topics such as scientific realism, explanation, and emergence: and it is indeed to these kinds of philosophical questions emerging from the study of dualities that we will turn in the next Chapter.

\section{Philosophical Questions}\label{philiss}

The philosophical literature on dualities has stressed how dualities relate to a wide range of philosophical topics: from general philosophical questions, such as scientific realism,\footnote{For discussions, especially in connection with under-determination arguments, see for example \citet{matsubara2013realism}, \citet{huggett2013emergent}, \citet{read2016interpretation} and \citet{le2018duality}.} 
scientific explanation and understanding,\footnote{See e.g.~\citet{de2018interpreting} and \citet{de2020precipice}.} 
to more specific questions concerning scientific theories that are mathematically well-developed, such as theoretical equivalence, the relation between symmetries and dualities, and the heuristic value of dualities.\footnote{See, among others, \citet{rickles2011interpretation}, \citet{teh2013holography}, \citet{dieks2015emergence}, \citet{de2016relation}, \citet{rickles2017dual}, \citet{de2017comparing}, \citet{castellani2017duality}, \citet{de2017spacetime}, \citet{huggett2017target}, \citet{butterfield2021dualities}, \citet{de2021theoretical}, and \citet{de2021symmetry}.}

In this Chapter, we will discuss three topics on which dualities and quasi-dualities bear. We begin with the general (perennial!) question {\it What is a scientific theory?} This is of course one of the defining questions of philosophy of science, since the early 20th century when the logical empiricists proposed a model of scientific theories that is now known as `the received view', or the syntactic conception, of scientific theories (\citet{putnam1966theories}). The received view came under attack in the second half of the 20th century, when the rival semantic conception of scientific theories was developed.\footnote{For a review of the criticisms, see \citeauthor{1974frederick} (1974:~pp.~63-72)).} 
Despite these attacks, it is now recognised by many authors that the syntactic conception retains many of its virtues, and a recent consensus has arisen that the syntactic and semantic conceptions do not after all differ as much as one might think. In fact, most authors agree that any reasonable formulation of a scientific theory requires both syntactic and semantic aspects.\footnote{For discussions, see for example \citeauthor{lutz2012straw} (2012:~p.~93), \citeauthor{lutz2017syntax} (2017:~pp.~345-347) and \citeauthor{frigg2022models} (2022:~p.~167).}

We will here first discuss the question {\it What is a scientific theory?}, from two different perspectives: the first is about {\it theory individuation}, and it is independent of one's preference for a syntactic or a semantic conception of theories. The second is a more specific modification, or further elaboration, of the {\it semantic conception} of theories. Thus Section \ref{TE} first discusses whether dualities can be taken as criteria of theory individuation: namely, can dual models be taken to be one and the same theory, i.e.~to be theoretically equivalent? Section \ref{QD} then discusses a modification of the semantic conception of theories that is suggested by quasi-dualities: namely, the `geometric view of theories'. 

The second question that this Section addresses is, in some sense, about `how to go beyond dualities with emergence'. The logic here is analogous to that for another inter-theoretic relation, viz.~reduction. Since reduction and emergence are, {\it prima facie}, incompatible, given a reduction of one theory to another, there is a natural question of whether emergence is possible. Likewise, duality also seems to exclude a relation of emergence. Nevertheless, in the physics literature, duality and emergence are often seen as two sides of the same coin (this is also how reduction and emergence are seen in most of the physics literature). Thus Section \ref{sec: emerg} will discuss the extent to which dualities are compatible with claims of emergence. Section \ref{sec: faq} ends with a number of related FAQs. 

Thus we will set aside other important philosophical questions, such as under-determination and understanding, where  dualities have been discussed (although we will occasionally mention scientific realism): for a discussion of these issues, see \citeauthor{deharobutterfieldOUP} (2025).

\subsection{Theoretical equivalence}\label{TE}

In this Section, we will briefly discuss one of the main questions that philosophers of dualities have focussed on in recent years: whether duals `say the same thing, in different words', so that they are mere reformulations of a single theory. Since this is an old and much-debated question in the philosophy of science, Section \ref{TEPS} first discusses two of the main criteria of theoretical equivalence that have been given in the literature, and it makes a first rough distinction of ways in which this notion applies to dualities. Section \ref{fice} then discusses two criteria that are jointly required for duals to be theoretically equivalent.

\subsubsection{Theoretical equivalence in philosophy of science and some interpretative options}\label{TEPS}

`Theoretical equivalence' is a broad term, which different authors have different ways of making precise. In general, `theoretical equivalence' is understood as `full equivalence' of (physical) theories. Theories that are theoretically equivalent `say the same thing, in different words': their disagreements are merely verbal, they are equivalent formulations, or `mere reformulations', of one and the same theory. 

In philosophy of science, two influential criteria that have been proposed as conditions for theoretical equivalence are due to \citeauthor{glymour1970theoretical} (1970:~p.~279) and \citeauthor{quine1975empirically} (1975:~p.~320).\footnote{See the exposition and comparison of these two views in \citet{barrett2016glymour}.} 
Glymour's inter-translatability criterion says that two theories are {\it definitionally equivalent} iff they have a common definitional extension, i.e.~if the signatures of each of these theories can be extended to include the symbols of the other theory. (`Signature' here means the non-logical vocabulary, e.g.~predicates, as against logical connectives like `and' and `or'.) If the thus extended theories are logically equivalent, then the original theories are definitionally equivalent, i.e.~they are inter-translatable. 

Quine, rather than extending the signatures of the two theories, maps one theory into the other by a {\it reconstrual} of the predicates of one theory in terms of the predicates of the other theory, such that the reconstrual of one theory is logically equivalent to the other theory. 

It is worth noting, and this will feed into our analysis of dualities below, that neither Glymour's nor Quine's criteria are (what is sometimes called) ``purely formal'' criteria, for two reasons. First, both criteria include a semantic component: the requirement that the extended theories (in Glymour's case) or the reconstrued theories (in Quine's case) are logically equivalent often involves considering their models. For, while first-order logic's completeness allows us to reformulate logical equivalence in terms of provability of sentences, the traditional philosophical interpretation includes the semantic aspect.\footnote{Beyond first-order logic, syntax and semantics can of course ``come apart''. For example, \citeauthor{van2001correspondence}'s (2001:~p.~342) bisimulation theorem gives a criterion for when a modal formula can be translated into a first-order formula. This, in effect, characterises when a fragment of first-order logic can have a translation from modal formulas. For a philosophical discussion, see \citeauthor{deharobutterfieldOUP} (2025:~pp.~378--381). For a discussion of expressive completeness, see \citeauthor{gradel2014freedoms} (2014:~pp.~3, 10). For the significance of this theorem and its generalizations, see \citeauthor{venema2014expressiveness} (2014: pp. 33–34).}

Second, Glymour is explicit that his criterion is only a necessary, but not a sufficient, condition for theoretical equivalence: and this is indeed how any such criteria have usually been interpreted in the philosophy of science literature.

As Chapter \ref{dualint} emphasised, we can think of a duality as a formal relation between models. Therefore, as our examples of dualities in the previous Chapter have illustrated, the interpretations of dual models can take a rich variety of forms. This richness is expressed by our labels `hard-easy', `elementary-composite', etc., for the contrasting  behaviour of duals. What does this imply for the question of the theoretical equivalence of duals? 

The answer is that the possibility of theoretical equivalence depends on various factors, which we will discuss in Section \ref{fice}: and so, that the verdicts differ by case. There are three broad types of possible cases, of which the first two are extremes, and the third one is an intermediate case that for us will be the most interesting one: (i) some cases where duals can clearly be, and are commonly, taken to be theoretically equivalent, (ii) some cases where duals are clearly not theoretically equivalent (for none of the systems they can possibly describe, i.e.~there is no system that is jointly described by both duals), and (iii) intermediate cases, where the possibility of theoretical equivalence requires further analysis. Let us note that, although cases (i) and (iii) have clear instantiations, it is not clear that there are any examples of type (ii). Below we will suggest an example that may be of this type.

(i) $\,$ In the first case, there is an internal interpretation (see the end of Section \ref{2.1}) on which dual models are standardly taken to be theoretically equivalent. For example, although models of quantum mechanics in the position and in the momentum representation look very different, once they are formulated in a basis-independent formalism, it is clear that they share a common core with an internal interpretation, according to which they use different variables to describe the same physical situations (see Section \ref{3.1}). 

(ii) $\,$ As an example of the second case, we can take the Kramers-Wannier high vs.~low temperature duality of the two-dimensional Ising model (see Section \ref{KWd}), as it is usually interpreted. On the standard way of interpreting the Ising model in statistical mechanics, an internal interpretation in terms of a single physical situation or theory does not seem to exist: since a lattice at high temperature is clearly not physically equivalent to a lattice at low temperature. In such cases, duals are not empirically equivalent, and a fortiori they are {\it not} theoretically equivalent.

(iii) $\,$ Most other cases we have discussed in our examples are of the third type, i.e.~they are in-between these two extremes, so that further analysis is required to decide whether, and under what conditions, there is theoretical equivalence. This includes examples of dualities where an internal interpretation is the most plausible interpretation, and examples where a fully  internal interpretation for a given type of system does not exist. 

For example, while it seems prima facie reasonable to say that electric and magnetic duals describe different physical situations, one can imagine a possible world where `purely electric' and `purely magnetic' states do not exist, and so where the duals are in fact theoretically equivalent (see the discussion at the end of Section \ref{EMMa}). 

But it is also clear that, in this kind of case, it takes some work to develop the internal interpretation, because it is not a priori clear that the common core theory has an ontology that corresponds to the kind of physical system that one aims at describing.\footnote{\citeauthor{read2020motivating} (2020:~pp.~266, 276) have called this position {\it motivationalism}.}

\subsubsection{A formal and an interpretative criterion of equivalence}\label{fice}

The upshot of the above discussion is that, although formal equivalence (in particular, duality) is a necessary condition for theoretical equivalence, it is not sufficient. Thus we require, in addition, an interpretative criterion. Summing up, theoretical equivalence requires two conditions:

{\bf (A)\hspace{.3cm}A formal criterion of equivalence:} namely, {\it duality}.

{\bf (B)\hspace{.3cm}An interpretative criterion of equivalence:} namely, duals have the {\it same domain of application}. More strongly, the requirement is that there is an internal interpretation, compatible with the duality.

Let us say a bit more about the motivation for adopting these two criteria, and about how the criteria are to be understood.

About (A): by `formal criterion', we mean a non-interpretative criterion of equivalence. Depending on how scientific theories are formulated, this criterion can be either syntactic or semantic, or both: as we discussed above, Glymour and Quine's criteria both have a syntactic and a semantic component, yet they are formal i.e.~non-interpretative in that, by themselves, they do not introduce additional requirements that the physical interpretations of the theories must satisfy (which is done by condition (B)).\footnote{Glymour's and Quine's criteria are not the only formal criteria in our sense that combine syntactic and semantic aspects. Others include: bisimulation, viz.~a relation between modal logic and a first-order language that preserves the accessibility relations and semantics; other generalizations of bisimulation, and categorical equivalence. For a discussion of some of these criteria, see \citeauthor{deharobutterfieldOUP} (2005:~Chapter 11).}

For physical theories, which are normally presented by mathematical physicists using set theory, the natural criterion of formal equivalence is the isomorphism criterion. For, as \citet{de2021theoretical} and \citeauthor{deharobutterfieldOUP} (2025) argue, some of the main critiques of the isomorphism criterion that have been given for physical theories stumble, because (i) either they rely on incorrect construals of the spaces that are required to be isomorphic; or (ii) they give incorrect treatments of the interpretative criteria, i.e.~of point (B).\footnote{See \citeauthor{deharobutterfieldOUP} (2025, Section 11.3).}

For physical theories as standardly formulated by mathematical physicists in set theory, physicists take duality, as an appropriate isomorphism between physical theories, to be {\it the} correct criterion of formal equivalence. We argue that the wealth of worked-out examples of significant dualities in physics justifies our endorsing this verdict.

Reformulating physical theories using other mathematical tools (e.g.~of formal logic, category theory, etc.) may allow for more mathematically precise criteria of formal equivalence: but it seems safe to say that any such improvements will be precisely that---{\it reformulations} of physical theories in a mathematical language that is more sophisticated. What we argue is that, {\it given} a physical theory formulated using set theory, as is standard in mathematical physics, duality is {\it the} correct formal criterion of equivalence.\footnote{For a detailed unpacking of the phrase `thus formulated', in terms of model roots that represent a common core that is preserved by the isomorphism, vs.~specific structure that is not preserved by the isomorphism, see \citeauthor{deharobutterfieldOUP} (2025:~Chapter 2). In Chapter 11, De Haro and Butterfield argue that the duality criterion, compared to other, logically weaker, criteria that have been proposed, has the right logical strength. They also argue that, unlike other criteria of theoretical equivalence that have been proposed, dualities satisfy two {\it principles of interpretation} that any scientific theory that is formulated in a satisfactory way, and thus any account of theoretical equivalence, should satisfy.}

About (B): the requirement is that equivalent models have the same physical semantics, i.e.~the same domain of application (i.e.~one requires numerical identity), and also that elements (i.e.~states and quantities) that are formally equivalent (i.e.~duals) are mapped into the same elements in the domain of application. In other words, the duality and the interpretation, both construed as particular kinds of maps, commute. 

We will discuss interpretation further below. Let us here make a comment about the relation between the requirements (A) and (B): the literature on theoretical equivalence has mostly focussed on (A), where various formal criteria have been investigated.\footnote{Some proponents of the formal criteria have considered an account where theoretical equivalence is a formal criterion (A), provided this formal criterion also respects empirical equivalence. But note that this is not what is meant by the mixed approach, i.e.~combining (A) and (B), since empirical equivalence is too weak a criterion of interpretative equivalence. In other words, formal equivalence, conjoined with empirical equivalence, is not a sufficient condition for theoretical equivalence, since theoretical equivalence is {\it full} equivalence, and includes interpretative equivalence.} 
When the literature has focussed on (B) in detail, this has sometimes been at the expense of (A). For example, \citet{coffey2014theoretical} has argued that judgements of theoretical equivalence boil down to judgements of interpretative equivalence: `claims of theoretical equivalence are normative claims about how theoretical formalisms ought to be interpreted' (p.~823). Formal considerations are relevant {\it only} in so far as they shape interpretative judgements. 

By contrast, the recent philosophical literature on dualities has argued that both possible extremes, i.e.~defining theoretical equivalence solely in terms of (A) or solely in terms of (B), are too weak, and that in general the only correct way to define theoretical equivalence is by using a mixed approach that requires {\it both} (A) and (B).\footnote{That (A) and (B) are both required for theoretical equivalence is by no means a new point. This was, for example, one of the central points of \citeauthor{glymour1977epistemology}'s (1977:~pp.~237, 242) `gorce plus morce' example, and his subsequent critique of the purported equivalence between Newtonian gravitation and geometrized Newtonian gravitation. However, this point seems to have been overlooked in some of the recent work on theoretical equivalence in philosophy of science. (Consequently, the point has been reiterated by \citeauthor{van2014one} (2014:~p.~278).) But it is also worth noting that the recent discussions of dualities have taken these arguments further: see e.g.~\cite{le2018duality}, \cite{read2020motivating}, \citet{butterfield2021dualities} and \cite{de2017spacetime,de2021theoretical}. It seems that this has, in part, been possible thanks to (i) the fact that, unlike examples like the different formulations of Newtonian gravitation, duals typically look very different, and so this requires rethinking the role of unobservables in relations of theoretical equivalence; (ii) the variety and physical salience of the examples, with rich and established interpretations in various areas of physics, which makes the examples vivid and striking, and also easier to conceptualise (as against pairs of theories that differ only subtly in their mathematical formulations).}

One way to argue for the need for a mixed approach is from what is usually required for a conception of a scientific theory and, more specifically, a physical theory, especially in theoretical physics. For, even though there are different (even conflicting) views of what a scientific theory is and how it is best formulated, major accounts such as the syntactic and the semantic conceptions of theories all require that scientific (more specifically, physical) theories have a formal or mathematical component and an interpretation (and, in particular, scientific statements have both a formal and an interpretative component). Thus for two theories to be equivalent, it is not sufficient to require that their interpretative components match: their formal components must also match. Therefore, a conception of the equivalence of theories must include both (A) and (B). (We will return to this topic in the FAQs in Section \ref{sec: faq}.)

\subsection{Dualities and the geometric view of theories}\label{QD}

Recall, from the preamble of this Section, our first overall question: {\it What is a scientific theory?} On the {\it semantic conception} of theories, the answer to this question is that a theory is {\it a collection of models}. These models are mathematical structures, usually defined directly using a set of elements that satisfy certain conditions (which typically include the satisfaction of the theory's defining equations). Thus a typical example of a physical theory on the semantic conception is: a collection of vector spaces, each equipped with a set of quantities (linear operators), a dynamics (in the Schr\"odinger picture, a deterministic rule for evolving an initial vector into a final vector), and a rule for evaluating quantities on states.

We will not here be concerned with the question of what type of language is required for the semantic conception, but rather with the question of whether this basic view, that `a theory is a collection of models', can describe the relevant examples from physics, or whether it is in need of modification. The basic view is often called a {\it flat view}, because the models are a collection, with no particular structure on them. In set-theoretic terms: a theory is a (bare) set of models, with no structure defined on that set.\footnote{For a formulation and critique of this view, see \citeauthor{halvorson2012scientific} (2012:~pp.~204-205).}

How do dual models fit this conception of a theory as a collection of models? In so far as dual models instantiate or represent a bare theory, with no further structure defined on the set of models, taking the set of duals (together with other models that may also be representations of the same bare theory) to be a theory seems to agree with a version of the semantic conception.\footnote{Also the defenders of the semantic conception agree that a conception of a scientific theory as `a collection of models' is too coarse, and is not what was originally meant by this conception, because the models are instantiations of a theory that is formulated in a mathematical language. For example, \citeauthor{glymour2013theoretical} (2013:~p.~289) has argued that, since a theory can be formulated in different languages, and the sentences of two such formulations can be inter-translated, the semantic conception requires that there are corresponding inter-relations between the models, or classes of models, of the theories. In other words, the set of models inherits inter-relations from its linguistically formulated theory. This agrees with our earlier statement, that the differences between the syntactic and semantic conceptions are smaller than it first seems. However, we are also going to argue that it is not enough to inherit just the structure from the inter-relations of the syntactically formulated theory.} 
For, just as in the semantic conception, each of the duals is itself a triple of state-space, set of quantities and dynamics.\footnote{However, one can see, also here, that the (simple gloss of the) semantic conception is insufficient. For the bare theory is itself also a structure, and bears a relation of homomorphism to its models, which is not taken into account by the simple gloss. \citeauthor{deharobutterfieldOUP} (2025) give an alternative syntactic reading, where the bare theory is a set of sentences, and the (dual) models are the structures that makes these sentences true. This formulation then fits better the model-theoretic view of theories than the simple views that are sometimes discussed in the philosophy of science.}

What we will argue in this Section is that this view of a physical theory, as a flat collection, or bare set, with no structure defined over the models, may only be valid in very special cases. Indeed, in general, we take a physical theory to be a {\it geometric object}: in our examples, it is a differentiable manifold.\footnote{In other examples, it is an algebraic variety. For a discussion, see \citeauthor{deharobutterfieldOUP} (2025). From a very different perspective, \citeauthor{halvorson2017categories} (2017:~pp.~411-412) have also advocated that the set of models is equipped with topological structure. In the context of classical mechanics, see \citeauthor{curiel2014classical} (2014:~pp.~275, 318); and, in the context of general relativity, see \citeauthor{fletcher2016similarity} (2016:~p.~366).}

To see this, recall, from Section \ref{sec: EMQFT}, that we can view the low-energy limit of supersymmetric Yang-Mills theory in terms of its {\it moduli space} and the quantities defined on it, i.e.~the space of configurations of the relevant fields that minimise the quantum potential (in the case of $\mathcal{N}=2$ supersymmetric Yang-Mills theory, the relevant field is the Higgs field). The information about the states is encoded not only in the location on the moduli space, but also in the topological and geometric structures defined on it (in our example, the moduli space is a two-dimensional connected, but not simply-connected, differentiable manifold, with a complex structure and a positive-definite K\"ahler metric). Thus the moduli space, together with the set of geometric quantities defined on it, defines the physical theory (here, at low energies). 

In more detail:---Requiring that a physical theory is a differentiable manifold of dimension $n$ means that the relevant state-space is a topological space that satisfies the axioms of a differentiable manifold. The  homeomorphisms from open subsets of the space onto $\mathbb{R}^n$ (or other model space like $\mathbb{C}^n$) are given by a set of quantities (at least $n$ of them) evaluated on the states that, together, form a set of coordinates: in a quantum theory, the value of a quantity is the expectation value of an operator evaluated on a given state. The open sets can be defined in various ways, but one natural choice that appears in the examples is to take as an open set the region of validity of a given formulation of the theory, in terms of a given set of quantities whose values are finite in that region. Indeed, since the coordinates are local homeomorphisms (where, by `local', we mean on the open set that is mapped by the homeomorphism), each coordinate is a continuous function and also has a continuous inverse. The smooth transition functions on the overlaps between the open sets are duality or quasi-duality maps. 

Thus from the perspective of the geometric view, more specifically of the requirement that a physical theory is a differentiable manifold, dualities and quasi-dualities can be understood as transition functions between coordinatizations of the manifold, i.e.~the physical theory, in overlapping regions. On a given overlap, we have different possible coordinatizations, i.e.~different formulations of the same (sub-)set of states and quantities, and these different formulations are related to each other by (quasi-)dualities.

The analogy with Kramers-Wannier duality for the Ising model in Section \ref{KWd} can help us better understand the physical significance of the different regions of the manifold and the coordinates on them. Recall the two phases of the Ising model, the low-temperature and high-temperature phases. At low temperatures, the magnetization takes a non-zero value and serves as an order parameter for this phase: the non-zero value indicates the broken symmetry of the phase. At the critical temperature, the magnetization drops to zero, and it is no longer a good order parameter above the critical temperature, where it remains zero. However, above the critical temperature there is a dual Ising model (see Eq.~\eq{KWdual}) with a {\it dual magnetization}, which {\it is} a good disorder parameter at high temperatures, since its non-zero value is indicative of the symmetric phase (which is the broken symmetry phase of the dual).

On this analogy, for a quantum field theory like supersymmetric Yang-Mills theory, a quantity that is a coordinate for a region of the manifold is like an order parameter in statistical mechanics. Its non-zero vacuum expectation value is indicative of a physical phase, in the sense that it has some characteristic type of macroscopic behaviour (for a more detailed discussion, see the second example in Section \ref{emdsym}, especially towards the end).

Let us briefly return to our motivating question, of how the `flat view' of theories is replaced by a `geometric view'. The flat view is the degenerate case in which there is no geometric or other structure on the set. But in general, physical theories do have at least topological and differentiable structure: and, we have argued, there is often also geometric structure that the flat view does not account for. These geometric structures contain information about the states and the quantities of the theory. In particular, the global structure can be used to distinguish different physical phases.

\subsection{Emergence}\label{sec: emerg}

An important philosophical topic arising from studies of dualities, which has received significant attention in the context of AdS-CFT, is emergence. In this Section, for ease of exposition, we will take AdS-CFT as our case study, and illustrate issues of emergence and duality in terms of it. However, the general morals that we draw, and more generally the issues that we highlight in this Section, apply across a wide range of dualities. Thus our focus on AdS-CFT both makes our discussion specific, and connects to the existing philosophical literature on duality and emergence.

There are at least two ways to think about the interaction between emergence and dualities, in particular AdS-CFT, depending on how we choose our emergence base, or comparison class, to which we compare the emergent entities or theory:\footnote{For more details about emergence, see \citet{palacios2022emergence}.} 
(i) emergence can be of the bulk from the boundary; (ii) the semi-classical bulk gravity theory emerges from fundamental quantum gravity data that is currently known only in terms of a dual CFT; (iii) there is no duality but only an effective duality. In all cases, we take the semi-classical bulk theory as emergent: then the question is whether it emerges from a fundamental bulk theory or from a fundamental boundary theory. We discuss each option in turn.

(i) {\it Boundary-to-bulk emergence with duality?} It seems natural to say that the boundary theory, whose exact definition is known, gives rise to the bulk theory, whose definition is, at best, known in perturbation theory (e.g.~because the bulk dual is a low-energy limit of string theory). This suggestion has been made in a variety of contexts, and it is indeed a natural way to think about AdS-CFT. For example, \citeauthor{horowitz2009gauge} (2009: p. 178) are prominent advocates of such a perspective on AdS-CFT-duality:
\begin{quote}
    AdS-CFT-duality is an example of emergent gravity, emergent spacetime, and emergent general coordinate invariance. But it is also an example of emergent strings! We should note that the terms ‘gauge-gravity duality’ and ‘gauge-string duality’ are often used ... to reflect these emergent properties.
\end{quote}

Despite its apparent intuitive appeal, the literature on dualities has argued (and we will endorse this view) that this is incorrect. For, as discussed in \citeauthor{dieks2015emergence} (2015), the claim of emergence that \citeauthor{horowitz2009gauge} (2009) make is incompatible with the claim that there is a duality between quantum gravity in AdS and the boundary CFT.\footnote{A similar argument is also made in \citeauthor{teh2013holography} (2013:~p.~310).} 
The reason is as follows. Duality is an isomorphism between models (in the sense of Chapter \ref{dualint}): thus it is a symmetric relation: if $a$ is dual to $b$, than $b$ is dual to $a$. 

The fact that dualities are symmetric relations is problematic for the claim of emergence because the latter is an \textit{asymmetric} relation, so that if $a$ emerges from $b$, then $b$ does not emerge from $a$ (i.e.~not in the same respects, or under the same conditions). We can illustrate this requirement in the prototypical examples, e.g.~emergence of thermodynamics from statistical mechanics, or emergence of the classical limit from a quantum theory. In the emergence of thermodynamics from statistical mechanics, we take thermodynamics to depend on statistical mechanics, while the converse is not true: and likewise for the emergence of the classical limit from the quantum theory, where the classical limit depends on the quantum theory, but not vice versa.

%Thus the reason why emergence of the bulk from the boundary is incompatible with the duality should be evident: if the relation between the bulk and the boundary is one of emergence, then it cannot be a symmetric relation, but rather it must be asymmetric. Also, u
Under the current assumption that the relation between the bulk and the boundary is a duality, AdS-CFT is a duality between a bulk theory (viz.~quantum gravity in AdS) and a boundary theory (viz.~the boundary CFT). And since a duality is a symmetric relation, the relation between the bulk and the boundary cannot be one of both duality and emergence, because this would imply the contradiction that this relation is both symmetric and asymmetric. Thus emergence between models is incompatible with their being duals. For a discussion of this argument, see \citet{dieks2015emergence} and \citet{de2017dualities}.\footnote{\citet{deharobutterfieldOUP} argue that, on grounds of the physical interest of the emergence relation, and in the way that these statements are usually made in the physics literature, the (putative!) duality and emergence relations must be the same relation (which is incompatible with the fact that one is asymmetric while the other is symmetric---hence `putative'). This implies that resolving the tension between emergence and duality by choosing them to be different relations is not an option.}

This said, there is important work that the notion of emergence can do for us in the context of dualities like AdS-CFT. For recall our discussion  of bulk reconstruction in Section \ref{sec: bulkads}, where we mentioned how entanglement wedge reconstruction is crucial to understand the encoding of bulk semiclassical gravity into the fundamental degrees of freedom, whose definition we know through the dual CFT. Since in this context we are moving from a more fundamental description, in terms of full quantum gravity, to a less fundamental one, in terms of bulk semiclassical gravity, it is natural to analyse this situation in terms of emergence. In particular, the map instantiating entanglement wedge reconstruction, as discussed in Section \ref{sec: bulkads}, would indeed encode the emergence relation between fundamental quantum gravity degrees of freedom and semiclassical gravity.\footnote{Slightly more precisely, it is the inverse of the encoding map, i.e.~the mapping from CFT quantities to local bulk semiclassical fields.}
This brings us to option (ii).

(ii) {\it Emergence beside duality.} We can avoid the arguments against case (i) by recalling our discussion of bulk reconstruction from Section \ref{sec: bulkads}. There, the relevant relation is between two different levels on one side of the duality, e.g.~the fundamental quantum gravity level and the semiclassical gravity level. Since this is a relation between models on one side of the duality, independent of the other side, it is compatible with duality. Namely, there is no reason to expect the relation between the models on one side of the duality to be symmetric: if anything, we would expect it to be asymmetric. (For example, the use of phrases such as `fundamental quantum gravity' builds in an asymmetry from the start.) Thus when discussing bulk reconstruction, we do not require the assumption that there is bulk-to-boundary emergence. Instead, we have an encoding relation between semi-classical gravity and the fundamental quantum gravity degrees of freedom; and since there is no obvious reason to regard {\it this} relation as symmetric, there is no obstacle to using emergence to analyse it. Thus we {\it can} have emergence beside duality, where the emergence and duality maps are along different directions, and relate different models.\footnote{For an accessible and physical argument for this view, see \citet{harlow2020black}.}

Thus in this type of emergence, there are two dual models at the lowest level, and two emergent dual models. Furthermore, there are duality maps relating the dual models, and emergence maps relating models at the lower and higher level, on the same side of the duality maps. This understanding of emergence is especially natural in bulk reconstruction in AdS-CFT because, as we discussed in Section \ref{sec: bulkads}, in bulk reconstruction we are interested in (a) the fundamental quantum gravity Hilbert {\it space of states}, understood as invariant between the two duals (and so part of the common core), but so far nonetheless only defined through the boundary CFT; and (b) {\it the set of quantities} defined in the emergent regime described by semiclassical gravity. Note that, just like the states, the quantities have boundary CFT duals, and so stand in an appropriate duality relation: these duals are indeed what bulk reconstruction seeks to identify. Furthermore, recall that these quantities are mapped to a code subspace of quantities on the CFT Hilbert space, as explained in \S\ref{sec: bulkads}; this is a subspace of the full CFT Hilbert space, which instead is supposed to encode the full, fundamental quantum gravity physics of holography. This restriction to a subspace naturally illustrates the idea that the semiclassical gravity description, confined to the code subspace, is emergent from the fundamental quantum gravity description. In this sense, bulk reconstruction shows that emergence between the fundamental quantum gravity level and the emergent semiclassical gravity level is a case of emergence-beside-duality, where both levels are described by holographically dual pairs.

(iii) {\it Effective duality.} If there is no duality, but only an effective duality, there is no objection to this relation also being emergence. For effective duality and emergence are both asymmetric relations. We have encountered effective dualities at various points in our discussion, especially in connection with quasi-dualities and the geometric view of theories. 
%This said, how should we analyze such a relation? There seem to be two main options here: either (1) we take the relation between gravity and the CFT to be a \textit{duality}, and hence to continue obtaining at all levels of fundamentality, or 
%(2) we do not, and hence expect the purported duality to be broken at some level; 

For ease of exposition, we will assume that: (1) the duality is effective in the sense that the low-energy models are duals, but the duality is broken at high energies, where one of the models is not well-defined. (2) The effective duality privileges the boundary CFT, in the sense that the CFT is well-defined at high energies, when the duality is broken and quantum gravity stops making sense.\footnote{Note that, as is customary in high-energy physics, we are here identifying levels of fundamentality with energy scales.} 

%These two perspectives give rise to two different ways to combine emergence and dualities, which we shall label as: (1) emergence-beside-duality and (2) effective duality.

%The first case, (1), is the one most naturally associated with the kind of scenario described in the context of bulk reconstruction: {\bf emergence-beside-duality} says that the emergence relation obtains together with the duality relation, i.e.~that both the fundamental and emergent theories stand in appropriate duality relations between the bulk and the boundary. Hence, In the second case, (2), there is an {\bf effective duality}, rather than a duality. 

%The point of effective duality in connection with emergence is that, if the duality only obtains in a certain regime, we are free to use the regime where the duality does not obtain to decide which dual is the most fundamental one and which one is judged to be emergent. In other words, an effective duality is itself an asymmetric relation, and as such does not exclude emergence in the way that dualities do. 

This notion fits some of the
%does not naturally fit examples like bulk reconstruction, and works better with 
examples in Section \ref{emdsym}, where we discussed effective dualities. A case where effective duality interestingly intersects with holography is the emergent gravity programme, as developed for example in \cite{verlinde2011origin,verlinde2017emergent}.\footnote{For a philosophical discussion of this scenario, see \citeauthor{dieks2015emergence} (2015:~Section 4).} 
Very roughly, the idea behind emergent gravity is to take a quantum system without gravity and without a gravitational holographic dual, often represented for simplicity as a collection of qubits (somewhat analogous to the boundary CFT of AdS-CFT), and study a limit of this system, often some type of thermodynamic limit, where the system develops emergent gravitational behaviour, in particular in the form of an effective holographic duality with some AdS spacetime. In this scenario, emergence goes from the fundamental, non-gravitational quantum system, to the emergent holographic description in the limit, which includes a gravitational theory. Since the emergence of gravity is mediated by the appearance of an effective holographic duality, it is natural to understand this case in terms of effective duality.

\subsection{FAQs about the philosophy of dualities}\label{sec: faq}

In this Section, we will answer five frequently asked questions about the philosophy of dualities.\\
\\
{\bf FAQ1. Is the common core ontology of two duals obtained by `deleting those variables or features that the duals do not share'?}\\
{\bf Answer:} No, it is not quite as simple as that. Building an ontology for the common core, i.e.~constructing an internal interpretation, may require a more sophisticated treatment of the features that are not shared by the duals than a simple ``deletion''. For example, electric and magnetic duals do not share `pure electric' and `pure magnetic' states: but this does not mean that the common core theory has no electric and magnetic variables and features at all, since it does have an {\it electromagnetic} field. We can use the analogy with the Lorentz transformations, which mix `purely electric' and `purely magnetic' fields (and forces) into an {\it electromagnetic field} that has both electric and  magnetic properties combined.\footnote{\citeauthor{deharobutterfieldOUP} (2025) distinguish two different ways in which a common core can be obtained: (i) by abstraction, and (ii) by augmentation.}

In the philosophical literature, this often goes under the name of `sophistication', and the standard example is gauge symmetries. If one wishes to interpret variables, such as gauge potentials related by a gauge transformation, as representing the same physical possibility, it is not necessary to use a reduced formalism that only trades in equivalence classes of gauge potentials. Instead, one can use a principal bundle, which keeps gauge-related gauge potentials, with gauge transformations as vertical automorphisms of the bundle. For a discussion of sophistication, see \cite{dewar2019sophistication,martens2021sophistry}.
(For more details, see the discussion at the end of Section \ref{EMMa}.)\\
\\
{\bf FAQ2. Why is condition (A) for theoretical equivalence (i.e.~having a formal criterion of equivalence, as in Section \ref{fice}) needed? Is condition (B) (i.e.~having an interpretative criterion) not enough?}\\
{\bf Answer:} No, condition (B) is not enough. This is because a theory in the natural sciences only says something if it uses an appropriate formalism: without a formalism, a physical theory of the usual type does not say (almost) anything, e.g.~it does not make any precise predictions. For example, the description and prediction of phenomena requires a formalism, and cannot be given in `purely interpretative' or `purely physical' terms. Reducing a theory to only its interpretative consequences, free of formalism, is hopeless, since it deprives one of the resources required to correctly express the concepts that are relevant to describe the envisaged domain of application. (The point here is not about using mathematics, as against some other language, in the theory's formalism. Rather, the point is that a formalism with sufficient expressive power, and a precise notion of formal equivalence, are required for theoretical equivalence: whether this formalism is written in a mathematical, or in a different, language.) Namely, the representing side of the interpretation relation determines the level of detail at which a theory can describe a domain of application.

As a consequence, it is a requirement that the mathematical structure of theoretically equivalent formulations should match, because this is required for an appropriate {\it ``inter-translation''}, so that the two theory formulations have the same level of detail and can ``say the same thing''. 

A maverick philosopher might claim that he or she can stipulate that the letter $E$ represents the theory of everything---on the grounds that, by a trivial semantic convention, any symbol can stand for anything: so that, in particular, $E$ stands for the (content of the) theory of everything. 

But this misunderstands what is required of a {\it scientific} theory: the symbol $E$, alone and by itself, even if endowed with an ``interpretation'' of some type, but free of mathematical formalism, is not a physical theory unless we are given the resources to make predictions and give explanations: namely, by adding a calculus, i.e.~a piece of mathematics. For the symbol $E$, without further syntactic rules, does not have the expressive power that enables the prediction of the cross-section of the collision between an electron and a positron, or of the half-life of carbon-14. In addition to interpretation, symbols require a calculus and rules of evaluation: which is what the formal aspects of scientific theories (as used in (A)) provide. (We also emphasised this aspect in Chapter \ref{dualint}, where we required that a physical theory is formulated as a triple of states, quantities and dynamics: \textit{some} such formulation is required for a physical theory.)

Thus, despite the idea of trivial semantic conventionality, formal inter-translation is required for theoretical equivalence, to secure that the two theories have the same predictive and descriptive power.\\
%While this is a question on which some philosophers may disagree, we have just argued that such philosophers can only do so by disregarding scientific standards about the predictive and descriptive power of scientific theories.\\
\\
{\bf FAQ3. Why is condition (B) for theoretical equivalence (i.e.~having an interpretative criterion of equivalence, as in Section \ref{fice}) needed? Is condition (A) (i.e.~having a formal criterion of equivalence) not enough?}\\
{\bf Answer:} No, condition (A) is not enough. This is because, except on a structural realist position, the semantics of a scientific theory is not only formal structure: it is not always, or not just, a model-theoretic semantics, with interpretations as models that are mathematical structures: even for the arch-anti-realist van Fraassen, the mathematical structure does not exhaust the semantics. 

This distinction, between model-theoretic semantics and the rest of the physical semantics, is of course shiftable and not absolute. Also, formalization helps to articulate and develop the physical interpretation.

But the distinction remains relevant within any given physical theory: see, for example, the hole argument, where given a certain type of spatiotemporal structure, there are significant disagreements about how this structure is to be interpreted, with what seem like various legitimate interpretative options.\\
\\
{\bf FAQ4. Surely two duals are automatically theoretically equivalent?} Some readers who are physicists might be surprised by our denying that duals are automatically theoretically equivalent (namely, whether they are theoretically equivalent is conditional on an explication of the common ontology). After all, one might argue, duality entails that any possible prediction, that one dual makes, can be ``translated'' into the corresponding prediction for the other dual, and vice versa. So, what else could duals possibly disagree on? It surely cannot be anything physical?

Another way of asking this question would be: setting aside those theories that, like the Ising model, only give partial descriptions of a physical system, how could duals that describe all the physical aspects of the system be inequivalent?\\
\\
{\bf Answer:} This is a version of FAQ3, and so one answer is `go to FAQ3': but since that was not meant as an exhaustive answer, we will here add some additional elements for a reply, which can be made at various levels.

About the `other way of asking this question': some authors have indeed required that one is justified in making a verdict of theoretical equivalence for duals that somehow ``describe the whole world''.\footnote{Various expressions of this requirement are in \citet{rickles2011interpretation}, \citet{dieks2015emergence}, \citet{de2017spacetime}, \citet{huggett2017target}, \citet{read2020motivating}, \citet{butterfield2021dualities}.}
This is a reasonable requirement, because theories that only describe partial systems might look equivalent on the partial systems, but might be inequivalent on the full system. 

For example, the electric-magnetic duality of the Maxwell equations in vacuum is an exact formal equivalence of the Maxwell theory and its dual, but expresses a symmetry only of regions of spacetime where there are no particles, but only the electromagnetic field. The duality is {\it not} valid in regions of space where there are electrical charges (the duality extends to such regions only if magnetic monopoles exist). \\
\\
{\bf FAQ5. Do dualities favour structural realism or anti-realism?}\\
{\bf Answer:} No, they do not, because they are compatible with realism about the common core theory (see FAQ1). 

However, it is worth noting that one could construe the common core strategy as being structuralist in nature. Structural realism, as introduced in philosophy of science by \cite{worrall1989structural}, is the belief that scientific theories tell us only about the form or structure of the unobservable world and not about its nature. It comes in two varieties \citep{ladyman1998structural}: an ontic one, where structure is all there is in the world, and an epistemic one, where structure is all we can come to know about the unobservable world. In both cases, it emphasizes the belief in the shared mathematical structure of theories and models, rather than in the specific objects to which they are ontologically committed. In the context of dualities, the common core theory characterises the shared mathematical structure between dual models. Therefore, the explicit construction of a shared structure can be seen as congenial to structural realism, which focusses on the preservation of mathematical relationships and structures across different theoretical frameworks. However, in order for this strategy to amount to a structural realist position, an additional step is required: namely, committing ourselves to the common core of duals while rejecting either that the nature of unobservable objects is correctly described by this common core (epistemic structural realism) or that there are unobservable objects at all (ontic structural realism).

Thus, while dualities do not intrinsically favour structural realism or anti-realism, the common core strategy, by emphasizing the importance of the shared mathematical structure, aligns with the principles of structural realism, especially in its `retention of shared structure'. 

Dualities do bear on the question of scientific realism along a different direction. Namely, as we discussed in Section \ref{fice}, the duality-based account of theoretical equivalence bears on the question of the structure and individuation of scientific theories. Thus this account highlights how dualities can shape our understanding of what constitutes a scientific theory and how theories are individuated based on their mathematical content and their interpretation.

\addtocontents{toc}{\protect\enlargethispage{2\baselineskip}}
\section{Conclusion}\label{conclusion}

This book has given an exposition of some of the main examples of dualities in physics, and of the philosophical and foundational questions that they raise. The Schema for dualities is a proposed  general framework for dualities that uses standard resources in mathematical physics. As our discussion has shown, using standard mathematical resources enables us to illustrate the Schema in both simple and advanced examples. Furthermore, various themes that run as a common thread through the examples are also illustrated.
%, and the themes that are associated with dualities, are instantiated, and cast light on, both simple and advanced examples from physics. 

Thus we began with simple examples, from Fourier duality in quantum mechanics to Kramers-Wannier duality in statistical mechanics and electric-magnetic duality in the Maxwell theory (Chapter \ref{3}). Then we went on to discuss more advanced examples (Chapter \ref{DQFTG}), from electric-magnetic duality in quantum field theory, to quantum gravitational examples like T-duality in string theory and the AdS-CFT correspondence, which provides a non-perturbative definition of quantum gravity for systems with AdS-like boundary conditions.

With these examples in mind, we discussed some of the philosophical work that has been done for dualities (Chapter \ref{philiss}), highlighting in particular the fundamental conceptual issues that they raise, and their intersection with various fundamental topics in the philosophy of physics and philosophy of science. Our focus has been directed mostly to theoretical equivalence, emergence, and the structure of scientific theories. All of these topics have received sustained attention in the philosophy of dualities, demonstrating the fruitful role of dualities in deepening our understanding of philosophical concepts, and also how philosophical reflection can help us understand the physical and conceptual structure of dualities, as in the example of emergence and gauge-gravity duality.

As we have seen, some of the main examples of dualities involve theories currently at the forefront of theoretical and experimental physics, and physical questions about the associated phenomena (e.g.~about the relation between particles and solitons, about the explanation of colour charge confinement, etc.) about which comparatively less work has been done in the philosophy of physics. Thus studying dualities promises to be a rich avenue for deepening our understanding of modern physics.

Much remains to be done, both in the physics and philosophy of dualities, to understand dualities and quasi-dualities, and explore the rich philosophical ground that they open up, and which the current literature has only begun to explore. This includes developing the common core for other dualities in fundamental physics. It also includes exploring the relationship between dualities and symmetries, and between dualities and various notions of fundamentality, which promise to provide interesting conceptual insights into both dualities and the foundations and metaphysics of physics. 

As for other questions for future work: it is important to gain a deeper understanding of how dualities involving quantum gravity and string theory, like holography and T-duality, relate to discussions about the emergence of spacetime, how spacetime emergence is compatible with dualities, and what notions of emergence are best suited to study these examples. Such an undertaking is sure to provide fertile ground for the interaction between fundamental physics and philosophy.
%, and to give new fascinating case studies and avenues for exploration to those interested in the philosophical foundations of quantum gravity and string theory.

Another topic that deserves significant attention, and which we have only begun to address here, is quasi-dualities and the geometric view of theories. Developing the geometric view in connection with theoretical equivalence, emergence and questions about fundamentality, seems to us to be of urgent importance. 

Indeed, the geometric view seems to be one of the most interesting ways in which dualities can help theorizing in both physics and philosophy. For it shows how reasoning about the structure of dualities can provide insights into new approaches to old problems in philosophy of science. And by developing these philosophical views, we can gain insights into the dualities that we started from, thus also improving our understanding of the physics. 

%Thus dualities are fertile ground to study the interaction between physics and philosophy, and how each can inform the other and improve our overall understanding of the physical world. We hope that this book has given the reader both a useful example of this way of doing philosophy, and inspired them and given them some tools to actively contribute to this project.

\section*{Acknowledgements}

We thank Jeremy Butterfield for comments on this manuscript. SDH thanks audiences at many graduate classes, seminars and conferences where this material was presented, especially at Urbino, Viterbo, Geneva, Munich, Amsterdam and Oxford. EC thanks audiences at graduate classes, and in Milan, Urbino, Amsterdam, and London, where this material was presented. We also thank two anonymous referees for their detailed comments. This project was funded by the Dutch National Science Agenda (Nationale Wetenschapsagenda, NWA) by NWO, under project number NWA.1418.22.029 `Is There Space and Time for Experimental Philosophy?', Small Innovative NWA projects. EC's research has also been supported by the SNSF project {\it Space, time, and causation in quantum gravity}.

\bibliography{refs}

\end{document}